\documentclass[10pt]{article}
\usepackage{a4wide,graphicx,float,latexsym,amssymb,subeqnarray,amsmath,amsfonts,epstopdf,color}

\def\R{\mathbb{R}}
\def\Z{\mathbb{Z}}

\newcommand{\lleft}{\rm{left}}
\newcommand{\rright}{\rm{right}}
\newcommand{\iin}{\rm{int}}

\newcommand{\rT}{{\rm{T}}}

\newcommand{\he}{{\widehat e}}
\newcommand{\hU}{{\widehat U}}
\newcommand{\hl}{{\widehat \lambda}}

\newcommand{\tY}{{\widetilde Y}}
\newcommand{\bU}{{\bar U}}

\title{Numerical study of splitting methods\\ for American option valuation}

\author{K.~J.~in 't Hout\footnote{Department of Mathematics and Computer Science,
University of Antwerp, Middelheimlaan 1, B-2020 Antwerp, Belgium.
\mbox{Email}: \texttt{\{karel.inthout,radoslav.valkov\}@uantwerpen.be}.}
~and R.~L.~Valkov\footnotemark[\value{footnote}]
}
\date{\today}

\begin{document}

\maketitle

\begin{abstract}
\noindent
This paper deals with the numerical approximation of American-style option values governed by partial differential complementarity problems.
For a variety of one- and two-asset American options we investigate by ample numerical experiments the temporal convergence behaviour of three modern splitting methods: the explicit payoff approach, the Ikonen--Toivanen approach and the Peaceman--Rachford method. 
In addition, the temporal accuracy of these splitting methods is compared to that of the penalty approach.
\end{abstract}

\section{Introduction}\label{intro}
American-style options are one of the most common instruments on the derivative markets and their valuation is of major interest to the financial industry.
In this paper we investigate by ample numerical experiments the accuracy and convergence of a collection of recent splitting methods that are employed in the numerical valuation of one- and two-asset American options.

Let $u(s_1,s_2,t)$ denote the fair value of a two-asset American-style option under the Black--Scholes framework if at $t$ time units before the given maturity time $T$ the underlying asset prices equal $s_1\ge 0$ and $s_2\ge 0$.
Let $\phi(s_1,s_2)$ denote the payoff of the option and define the spatial differential operator
\begin{equation}\label{calA}
\mathcal{A}   =
\tfrac{1}{2} \sigma_1^2 s_1^2 \frac{\partial^2 }{\partial s_1^2}
+ \rho \sigma_1 \sigma_2 s_1 s_2 \frac{\partial^2 }{\partial s_1 \partial s_2}
+\tfrac{1}{2} \sigma_2^2 s_2^2 \frac{\partial^2 }{\partial s_2^2}
+ r s_1 \frac{\partial }{\partial s_1}
+ r s_2 \frac{\partial }{\partial s_2}
- r.
\end{equation}
Here constant $r$ is the risk-free interest rate, the positive constant $\sigma_i$ denotes the volatility of the price of asset $i$ for $i=1,2$ and the constant $\rho\in [-1,1]$ stands for the correlation factor pertinent to the two underlying asset price processes.

It is well-known that the function $u$\, satisfies the partial differential complementarity problem (PDCP)
\begin{equation}\label{PDCP}
 u\ge \phi,
\quad \frac{\partial u}{\partial t}\ge \mathcal{A} u,
\quad (u-\phi)\left(\frac{\partial u}{\partial t} -
\mathcal{A} u\right)=0,
\end{equation}
valid pointwise for $(s_1,s_2,t)$ with $s_1> 0$, \,$s_2> 0$,\, $0< t \le T$.
The initial condition is prescribed by the payoff,
\begin{equation}\label{IC}
u(s_1,s_2,0) = \phi (s_1,s_2)
\end{equation}
for $s_1\ge 0$, \,$s_2\ge 0$ and boundary conditions are given by imposing \eqref{PDCP} for $s_1=0$ and $s_2=0$, respectively.
The three conditions in (\ref{PDCP}) naturally induce a decomposition of the $(s_1,s_2,t)$-domain: the early exercise region is the set of all points $(s_1,s_2,t)$ where $u=\phi$ holds and the continuation region is the set of all points $(s_1,s_2,t)$ where $\partial u/\partial t= \mathcal{A}u$ holds.
The joint boundary of these two regions is referred to as the early exercise boundary or free boundary.

For most American-style options both the option value function and the early exercise boundary are unknown in (semi-)closed analytical form.
Accordingly, one resorts to numerical methods for their approximation.
Following the method of lines, one first discretizes the PDCP \eqref{PDCP} in the spatial variables $(s_1,s_2)$ and next discretizes in the temporal variable $t$.
This leads to a linear complementarity problem (LCP) in each time step.

Various approaches have been proposed in the literature to handle these LCPs.
In this paper we study three modern splitting methods: the explicit payoff (EP) approach, the Ikonen--Toivanen (IT) approach and the Peaceman--Rachford (PR) method. 
The IT splitting approach was introduced in \cite{KR_IT04,KR_IT09} and has recently been combined in \cite{KR_HH15} with Alternating Direction Implicit (ADI) schemes for the temporal discretization.
The PR method was proposed in \cite{KR_LM79}.

At present the convergence theory pertinent to the IT splitting approach appears to be limited in the literature.
The main goal of this paper is to gain more insight into its convergence behaviour by numerically studying temporal discretization errors.
In addition, all splitting approaches above are compared to the popular penalty (P) approach, which was introduced for American option valuation in \cite{KR_FV02,KR_ZFV98a,KR_ZFV01}.
An outline of our paper is as follows.

In Section~\ref{spatial_discr} a suitable spatial discretization of the PDCP \eqref{PDCP} is formulated.
In Section~\ref{temporal} the temporal discretization methods under consideration are described: the \mbox{$\theta$-EP} method, the \mbox{$\theta$-IT} method, the PR method, the \mbox{$\theta$-P} method and three families of \mbox{ADI-IT} methods.
Section~\ref{interpret} provides an illuminating interpretation of the IT approach and the PR method.
Subsequently, an ample numerical study of the temporal discretization errors for all methods above is performed in Section~\ref{numerical_study}, where five American-style options are considered.
Conclusions are presented in Section~\ref{conclusions}.

\section{Spatial discretization}\label{spatial_discr}
The numerical solution of the PDCP (\ref{PDCP}) commences with the discretization of the spatial differential operator $\mathcal{A}$ defined by (\ref{calA}).
To this purpose, the unbounded spatial domain is first truncated to a square $(0,S_{\rm max})\times (0,S_{\rm max})$ with given value $S_{\rm max}$ chosen sufficiently large.
We prescribe homogeneous Neumann conditions at the far-field boundaries $s_1=S_{\rm max}$ and $s_2=S_{\rm max}$, consistent with the payoffs under consideration. 

For the spatial discretization a suitable nonuniform Cartesian grid is taken,
\begin{equation*}\label{grid2}
(s_{1,i}\,,\,s_{2,j}) \quad {\rm for}~ 0\le i\le m_1,\, 0\le j\le m_2,
\end{equation*}
where $0 = s_{k,0} < s_{k,1} < \ldots < s_{k,m_k} = S_{\rm max}$ is the mesh in the $k$-th spatial direction for $k=1,2$.
The use of nonuniform grids, instead of uniform ones, can yield a substantial improvement in efficiency.
We consider here the type of grid used in \cite{KR_HH12,KR_HH15}.
Let $k\in\{1,2\}$ and let $[S_{\lleft},S_{\rright}]$ be any given fixed subinterval of $[0,S_{\max}]$ that is of practical interest in the $k$-th direction.
Let parameter $d>0$ and let equidistant points \mbox{$\xi_{\min}=\xi_0 < \xi_1 < \ldots < \xi_{m_k}=\xi_{\max}$}
be given with
\begin{align*}
\xi_{\min} &= \sinh^{-1}\left( \frac{- S_{\lleft}}{d} \right),\\
\xi_{\iin} &= \frac{S_{\rright}-S_{\lleft}}{d},\\
\xi_{\max} &= \xi_{\iin} + \sinh^{-1}\left( \frac{S_{\max} - S_{\rright}}{d} \right).
\end{align*}
Note that $\xi_{\min} < 0 < \xi_{\iin} < \xi_{\max}$.
The mesh $0 = s_{k,0} < s_{k,1} < \ldots < s_{k,m_k} = S_{\rm max}$ is then defined through the transformation
\begin{equation*}
s_{k,l} = \varphi(\xi_l) \quad (0\le l \le m_k),
\end{equation*}
where
\begin{equation*}
\varphi(\xi) =
\begin{cases}
 S_{\lleft} + d\sinh(\xi) & ({\rm for}~\xi_{\min} \leq \xi \le 0),\\
 S_{\lleft} + d\xi & ({\rm for}~0 < \xi \leq \xi_{\iin} ),\\
 S_{\rright} + d\sinh(\xi-\xi_{\iin}) & ({\rm for}~\xi_{\iin} < \xi \leq \xi_{\max}).
\end{cases}
\end{equation*}
By construction, this mesh is uniform inside the interval $[S_{\lleft},S_{\rright}]$ and nonuniform outside, where the mesh widths outside the interval are always larger than the mesh width inside. 
The parameter $d$ controls the fraction of mesh points that lie inside.
Let $\Delta \xi = \xi_1-\xi_0$. 
It is readily shown that the above mesh is smooth, in the sense that there exist real constants $C_0, C_1, C_2 >0$ such that the mesh widths $h_{k,l} = s_{k,l} - s_{k,l-1}$  satisfy
\begin{equation*}\label{smooth}
C_0\, \Delta \xi \le h_{k,l} \le C_1\, \Delta \xi ~~ {\rm and} ~~
|h_{k,l+1}-h_{k,l}| \le C_2 \left( \Delta \xi \right)^2 ~~
{\rm uniformly~in}~\, l,\, m_k.
\end{equation*}

For our applications it turns out to be beneficial for accuracy if the given strike price $K>0$ of an option is located exactly midway between two successive mesh points.
This can be achieved with the above type of mesh as follows.
Fix an interval $[S_{\lleft},S_{\rright}]$ such that $K$ lies in the middle.
Then $K=\varphi(\xi_{\iin}/2)$.
Let $\nu\ge 1$ be any given integer and take
\[
\Delta \xi = \frac{\xi_{\iin}-2\,\xi_{\min}}{\nu},
\]
so that $\xi_{\min}=\xi_0 < \xi_1 < \ldots < \xi_\nu=\xi_{\iin}-\xi_{\min}$.
It holds that $s_{k,\nu} = \varphi(\xi_{\iin}-\xi_{\min}) = S_{\lleft} + S_{\rright}$.
The point $\xi_{\iin}/2$ is the middle of the interval $[\xi_{\min},\xi_{\iin}-\xi_{\min}]$ and lies exactly midway between two successive $\xi$-mesh points whenever $\nu$ is~odd.
This implies that $K$ lies exactly midway between two successive $s_k$-mesh points whenever $\nu$ is~odd.
Then let $m_k = m_k(\nu)$ be the smallest integer such that $m_k \Delta \xi \ge \xi_{\max} - \xi_{\min}$ and reset $\xi_{\max}$ to $\xi_{\min} + m_k \Delta \xi$.

The discretization of the operator $\mathcal{A}$ is performed using finite differences.
Let $f\!:\R\rightarrow\R$ be any given smooth function. 
Let $\{ x_l \}_{l\in \Z}$  be any given increasing sequence of mesh points and $\Delta x_l = x_l-x_{l-1}$ for all $l$.
For approximating the first and second derivatives of $f$ we consider the following well-known finite difference formulas:
\begin{subequations}
\begin{align}
f'(x_l) ~~\approx~~
&\alpha_{0} f(x_{l}) + \alpha_{1} f(x_{l+1}),\label{forw}\\
\nonumber\\
f'(x_l) ~~\approx~~
&\beta_{-1} f(x_{l-1}) + \beta_{0} f(x_l) + \beta_{1} f(x_{l+1}),\label{MethodB}\\
\nonumber\\
f''(x_l) ~~\approx~~
& \delta_{-1} f(x_{l-1}) + \delta_{0} f(x_l) + \delta_{1} f(x_{l+1})\label{diffusion}
\end{align}
\end{subequations}
with
\begin{align*}
&
&\alpha_{0}&= \tfrac{-1}{\Delta x_{l+1}},
&\alpha_{1}&= \tfrac{1}{\Delta x_{l+1}},\\
\beta_{-1}&= \tfrac{-\Delta x_{l+1}}{\Delta x_l (\Delta x_l + \Delta x_{l+1})},
&\beta_{0}&= \tfrac{\Delta x_{l+1}-\Delta x_l}{\Delta x_l \Delta x_{l+1}},
&\beta_{1}&= \tfrac{\Delta x_l}{\Delta x_{l+1} (\Delta x_l + \Delta x_{l+1})},\\
\delta_{-1}&= \tfrac{2}{\Delta x_l (\Delta x_l + \Delta x_{l+1})},
&\delta_{0}&= \tfrac{-2}{\Delta x_l \Delta x_{l+1}},
&\delta_{1}&= \tfrac{2}{\Delta x_{l+1} (\Delta x_l + \Delta x_{l+1})}.
\end{align*}
The approximation \eqref{forw} is the first-order forward formula.
The approximations \eqref{MethodB}, \eqref{diffusion}~are both central formulas that are second-order whenever the mesh is smooth.

For the discretization of the terms $\partial^2 u/\partial s_k^2$ ($k=1,2$) in $\mathcal{A}u$, formula \eqref{diffusion} is taken.
For the terms $\partial u/\partial s_k$ ($k=1,2$) formula \eqref{MethodB} is applied at each mesh point $s_{k,l}$ such that the corresponding coefficient 
\[
r s_{k,l} \beta_{-1} + \tfrac{1}{2} \sigma_k^2 s_{k,l}^2 \delta_{-1}
\] 
is nonnegative, otherwise formula \eqref{forw} is used.
This mixed central/forward discretization of the convection terms is often employed in the literature.
For the cross derivative term $\partial^2 u/\partial s_1\partial s_2$ the finite difference formulas used for $\partial u/\partial s_1$, $\partial u/\partial s_2$ are successively applied.
Concerning the boundary of the spatial domain, at $s_k=0$ all spatial derivative terms involving the \mbox{$k$-th} direction vanish, so that this part of the boundary is trivially dealt with ($k=1,2$).
At $s_k=S_{\rm max}$ the Neumann condition directly yields $\partial u/\partial s_k$ and $\partial^2 u/\partial s_k^2$ is approximated using a virtual point $s_{k,{m_k}+1} > S_{\max}$, where the value at this point is defined by linear extrapolation ($k=1,2$).

The given spatial discretization leads to a semidiscrete PDCP system 
\begin{equation}\label{lcp_ode}
U(t) \ge U_0,
\quad U'(t) \ge AU(t),
\quad (U(t) - U_0)^\rT (U'(t) - AU(t))=0
\end{equation}
for $0 < t \le T$ with $U(0)=U_0$.
Here $U(t)$ denotes the $M\times 1$ vector representing the semidiscrete approximation to the option value function $u(\cdot,\cdot,t)$ on the spatial grid, where $M=(m_1+1)(m_2+1)$. 
The $M\times M$ matrix $A$ and the initial $M\times 1$ vector $U_0$ are given, where the latter represents the payoff function $\phi$ on the spatial grid.
The vector inequalities are to be interpreted componentwise and the symbol $^{\rm T}$ denotes taking the transpose.

Taking into account the selection of the finite difference formulas and the boundary conditions, it readily follows that $-A$ is always an M-matrix
whenever the correlation $\rho=0$.
This feature is widely used~in the computational finance literature to prove favourable properties of numerical methods.
If $\rho\not=0$, then $-A$ is usually not an M-matrix anymore when standard finite difference formulas for the mixed derivative are applied.
Advanced techniques exist to overcome this, but in the present paper we shall adhere to the standard discretization above.

\section{Temporal discretization}\label{temporal}
For the temporal discretization of the obtained semidiscrete PDCP system \eqref{lcp_ode} we deal in Section \ref{thetamethods} with the well-known family of $\theta$-methods.
Special instances of this are the Crank--Nicolson (CN) method for $\theta=\tfrac{1}{2}$, and the backward Euler (BE) method for $\theta=1$.
Next, in Section \ref{ADIschemes} three prominent families of Alternating Direction Implicit (ADI) schemes are considered.

\subsection{$\theta$-methods}\label{thetamethods}
Let parameter $\theta>0$.
Let $I$ denote the $M\times M$ identity matrix.
Let $\Delta t = T/N$ with integer~$N\ge 1$ be a given time step and let temporal grid points $t_n = n\Delta t$ for integers $0\le n \le N$.
The $\theta$-method applied to \eqref{lcp_ode} defines approximations $U_n \approx U(t_n)$ successively for $n=1,2,\ldots,N$ by
\begin{subeqnarray}\label{fullPDCP}
&&U_n \ge U_0,\\
\nonumber\\
&&(I-\theta\Delta t A)U_n \ge (I+(1-\theta)\Delta t A)U_{n-1},\\
\nonumber\\
&&(U_n - U_0)^{\rm T} ((I-\theta\Delta t A)U_n - (I+(1-\theta)\Delta t A)U_{n-1})=0.
\quad\quad\quad
\end{subeqnarray}
\noindent
The fully discrete PDCP \eqref{fullPDCP} forms a linear complementarity problem (LCP) for the vector $U_n$.
Much attention has been paid in the literature to the solution of LCPs.
We consider in the following several approximation approaches that are popular in the present time-dependent context.

The explicit payoff (EP) approach is arguably the most commonly used in financial~practice.
It yields the simple method \eqref{LCP_explicit}, generating for $n=1,2,\ldots,N$ approximations \mbox{$\hU_n \approx U(t_n)$}.
\\\\
\noindent
{\it  $\theta$-EP method}\,:
\begin{subeqnarray}\label{LCP_explicit}
&&(I-\theta\Delta t A)\bU_n = (I+(1-\theta)\Delta t A) \hU_{n-1},\\
\nonumber\\
&&\hU_n = \max \{\bU_n\,,\, U_0 \}
\end{subeqnarray}
with $\hU_0 = U_0$ where the maximum of any two vectors is to be taken componentwise.
Method \eqref{LCP_explicit} can be regarded as a fractional step splitting technique in which one first performs a time step by ignoring the American constraint and next applies this constraint explicitly, compare~\cite{KR_BDR95}.
More precisely, the latter means projecting $\bU_n$ onto the closed convex subspace of vectors $V\in \R^M$ satisfying $V\ge U_0$. 
The computational cost per time step of the $\theta$-EP method is essentially the same as that in the case of the European counterpart of the option, which is very favourable.

The Ikonen--Toivanen (IT) operator splitting approach \cite{KR_IT04,KR_IT09} has the same computational cost.
It yields the
\\\\
\noindent
{\it $\theta$-IT method}\,:
\begin{subeqnarray}\label{LCP_IT}
&&(I-\theta\Delta t A)\bU_n = (I+(1-\theta)\Delta t A) \hU_{n-1}
+ \Delta t\,\hl_{n-1},\\\nonumber\\
&&\left\{\begin{array}{l}
\hU_n-\bU_n-\Delta t\,(\hl_n-\hl_{n-1})=0,\\\\
\hU_n\geq U_0,~~ \hl_n\geq 0,~~ (\hU_n-U_0)^{\rm T}\, \hl_n =0
\end{array}\right.
\end{subeqnarray}
with $\hl_0=0$.
The vector $\hU_n$ and the auxiliary vector $\hl_n$ are computed in two stages.
In the first stage, an intermediate approximation $\Bar{U}_n$ is defined by the linear system (\ref{LCP_IT}a).
In the second stage, $\Bar{U}_n$ and $\hl_{n-1}$ are updated to $\hU_n$ and $\hl_n$ by (\ref{LCP_IT}b).
It is easily seen that for these updates one has the simple formula
\begin{equation}\label{LCP_IT_update}
\left\{\begin{array}{l}
\hU_n = \max\left\{\bU_n-\Delta t\, \hl_{n-1}\,,\,U_0\right\},\\
\hl_{n} = \max\left\{0\,,\,\hl_{n-1}+(U_0-\bU_n)/\Delta t\right\}.
\end{array}\right.
\end{equation}

A related approach has been considered by Lions \& Mercier \cite{KR_LM79}, which was inspired by the original Peaceman--Rachford (PR) directional splitting scheme \cite{KR_PR55}.
It can be formulated as the
\\\\
\noindent
{\it PR method}\,:
\begin{subeqnarray}\label{LCP_PR}
&&(I-\tfrac{1}{2}\Delta t A)\bU_n = \hU_{n-1} + \tfrac{1}{2}\Delta t\,\hl_{n-1},\\\nonumber\\
&&\left\{\begin{array}{l}
\hU_n = \max\left\{(I+\tfrac{1}{2}\Delta t A) \bU_n\,,\,U_0\right\},\phantom{\left\{\hl_{n-1}\right\}}\\
\hl_{n} = \max\left\{0\,,\,U_0-(I+\tfrac{1}{2}\Delta t A) \bU_n\right\}/(\tfrac{1}{2}\Delta t).
\end{array}\right.
\end{subeqnarray}
A useful interpretation of the IT splitting approach and the PR method shall be given in Section~\ref{interpret}.

Let ${\it Large} >0$ be any fixed large number and integer $\kappa\ge 1$. 
The penalty approach has been proposed for American option valuation in \cite{KR_FV02,KR_ZFV98a,KR_ZFV01}.
It yields
\\\\
\noindent
{\it  $\theta$-P method}\,:
\begin{eqnarray}\label{LCP_penalty}
&&\left( I-\theta\Delta t A +P^{(k)}_n \right) \Bar{U}_{n}^{(k+1)} = (I+(1-\theta)\Delta t A) \hU_{n-1} + P^{(k)}_n U_0 \qquad\nonumber\\
&& {\rm for}~ k=0,1,\ldots,\kappa-1~~{\rm and}~~\hU_{n} = \Bar{U}_{n}^{(\kappa)}.
\end{eqnarray}
This forms an iteration in each time step.
Here $\Bar{U}_{n}^{(0)} = \hU_{n-1}$ and $P^{(k)}_n$ (for $0\le k < \kappa$) is~defined as the diagonal matrix with $l$-th diagonal entry equal to ${\it Large}$ whenever $\Bar{U}_{n,l}^{(k)} < U_{0,l}$ and zero otherwise.
In each time step, $\kappa$ linear systems have to be solved, involving different matrices.
Accordingly, the penalty method is computationally more expensive per time step than the three foregoing methods.
A natural convergence criterion is
\begin{equation}\label{conv_crit}
\max_l \frac{| \Bar{U}_{n,l}^{(k+1)}-\Bar{U}_{n,l}^{(k)}|}{\max \{1,| \Bar{U}_{n,l}^{(k+1)} | \} } < {\it tol} \quad {\rm or} \quad P^{(k+1)}_n = P^{(k)}_n,
\end{equation}
with given sufficiently small tolerance ${\it tol} >0$.
Let $\epsilon$ denote the machine precision of the~computer. 
A rule of thumb\footnote{Suggested to the authors by P.~A.~Forsyth.} for the choice of penalty factor is then
\begin{equation}\label{penalty fac}
{\it Large} \approx \alpha \frac{{\it tol}}{\epsilon}\quad {\rm with}~~\alpha = 10^{-2}.
\end{equation}
We have $\epsilon \approx 10^{-16}$ and choose ${\it tol} = 10^{-7}$ and ${\it Large} = 10^{7}$.
In our applications, the average number of iterations $\kappa$ per time step lies between 1~and~2.

\subsection{ADI schemes}\label{ADIschemes}
ADI schemes are attractive for the temporal discretization of semidiscrete multidimensional PDEs as their computational cost per time step is directly proportional to the number of spatial grid points $M$ from the semidiscretization, which is optimal.
For these schemes, the matrix $A$ is split into
\begin{equation}\label{splitA}
    A = A_0+A_1+A_2.
\end{equation}
Here $A_0$ is the part of $A$ that corresponds to the semidiscretization of the mixed derivative term.
This matrix is nonzero whenever the correlation $\rho$ is nonzero.
Next, $A_1$ and $A_2$ are the parts of $A$ that correspond to the semidiscretization of all spatial derivative terms in the $s_1$- and $s_2$-directions, respectively, and also contain an equal part of $-rI$.
These two matrices are essentially tridiagonal (that is, up to a possible permutation).

In the literature on the numerical valuation of European-style options, three prominent families of ADI schemes have been considered: the Douglas (Do) scheme, the Modified Craig--Sneyd (MCS) scheme and the Hundsdorfer--Verwer (HV) scheme, compare~e.g.~\cite{KR_IHF10}.
In \cite{KR_HH15,KR_HIHV10} these schemes have been adapted for the numerical valuation of American-style options by combining them with the IT splitting approach, leading to so-called \mbox{ADI-IT} methods:
\\\\
\noindent
{\it \mbox{Do-IT} method}\,:
\vspace{0.2cm}
\begin{equation}\label{Do}
\begin{cases}
Y_0 = (I+\Delta t A)\hU_{n-1} + \Delta t\, \hl_{n-1},\\
Y_j = Y_{j-1}+\theta \Delta t A_j \left(Y_j-\hU_{n-1}\right)~~~~~~~~~~~~~~~~~~~~~~~(j=1,2),\\
\Bar{U}_n = Y_2,\\\\
\hU_n = \max\left\{\Bar{U}_n- \Delta t\, \hl_{n-1}\,,\,U_0\right\},\\
\hl_{n} = \max\left\{0\,,\,\hl_{n-1}+(U_0-\Bar{U}_n)/\Delta t \right\}.
\end{cases}
\end{equation}
\\\\
\noindent
{\it \mbox{MCS-IT} method}\,:
\vspace{0.2cm}
\begin{equation}\label{MCS}
\begin{cases}
Y_0 = (I+\Delta t A)\hU_{n-1} + \Delta t\, \hl_{n-1},\\
Y_j = Y_{j-1}+\theta \Delta t A_j \left(Y_j-\hU_{n-1}\right)~~~~~~~~~~~~~~~~~~~~~~~(j=1,2),\\
\tY_0 = Y_0+\left( \theta \Delta t\, A_0 +(\tfrac{1}{2}-\theta) \Delta t A \right)\left(Y_2-\hU_{n-1}\right),\\
\tY_j = \tY_{j-1}+\theta \Delta t A_j \left(\tY_j-\hU_{n-1}\right)~~~~~~~~~~~~~~~~~~~~~~~(j=1,2),\\
\Bar{U}_n = \tY_2,\\\\
\hU_n = \max\left\{\Bar{U}_n- \Delta t\, \hl_{n-1}\,,\,U_0\right\},\\
\hl_{n} = \max\left\{0\,,\,\hl_{n-1}+(U_0-\Bar{U}_n)/\Delta t \right\}.
\end{cases}
\end{equation}

\clearpage
\noindent
{\it \mbox{HV-IT} method}\,:
\vspace{0.2cm}
\begin{equation}\label{HV}
\begin{cases}
Y_0 = (I+\Delta t A)\hU_{n-1} + \Delta t\, \hl_{n-1},\\
Y_j = Y_{j-1}+\theta \Delta t A_j \left(Y_j-\hU_{n-1}\right)~~~~~~~~~~~~~~~~~~~~~~~(j=1,2),\\
\tY_0 = Y_0+ \tfrac{1}{2} \Delta t A \left(Y_2-\hU_{n-1}\right),\\
\tY_j = \tY_{j-1}+\theta \Delta t A_j \left(\tY_j-Y_2\right)~~~~~~~~~~~~~~~~~~~~~~~~~~~(j=1,2),\\
\Bar{U}_n = \tY_2,\\\\
\hU_n = \max\left\{\Bar{U}_n- \Delta t\, \hl_{n-1}\,,\,U_0\right\},\\
\hl_{n} = \max\left\{0\,,\,\hl_{n-1}+(U_0-\Bar{U}_n)/\Delta t \right\}.
\end{cases}
\end{equation}
\vskip0.5cm
\noindent
The \mbox{Do-IT} method constitutes the basic \mbox{ADI-IT} method.
The \mbox{MCS-IT} and \mbox{HV-IT} methods form different extensions to this method, which require about twice the amount of computational work per time step.

In the \mbox{ADI-IT} methods above, the $A_0$ part is always treated in an explicit manner and the $A_1$~and~$A_2$ parts in an implicit manner.
In each time step, linear systems have to be solved with the two matrices $I-\theta \Delta t A_j$ for $j=1,2$.
Since these matrices are both tridiagonal, the solution can be done very efficiently by computing once, upfront, their $LU$ factorizations and then employ these in all time steps.
It thus follows that the computational cost per time step of each \mbox{ADI-IT} method is directly proportional to the number of spatial grid points $M$, which is very favourable.

Concerning the underlying ADI schemes it holds that the MCS and HV schemes both have a classical order of consistency (that is, for fixed nonstiff ODE systems) equal to two for any value~$\theta$.
We mention that the MCS scheme with $\theta=\tfrac{1}{2}$ is the so-called Craig--Sneyd (CS) scheme.
For the Do scheme, if $A_0$ is nonzero, then the classical order of consistency is only equal to one. 
This lower order is due to the fact that in this scheme the $A_0$ part is treated in a simple, forward Euler fashion.

\section{An interpretation of the IT approach and the PR method}\label{interpret}
In this section we present an illuminating interpretation of the IT splitting approach and the PR method. 
It is obtained upon rewriting the semidiscrete PDCP \eqref{lcp_ode} by means of an auxiliary variable $\lambda(t)$, often called a Lagrange multiplier:
\begin{subeqnarray}\label{lcp_ode_2}
&&U'(t) = AU(t)+\lambda(t),\\
\nonumber\\
&&U(t) \ge U_0,~~ \lambda(t)\ge 0,~~ (U(t) - U_0)^\rT \lambda(t)=0.
\end{subeqnarray}
Suppose for the moment that $\lambda(t)$ is known and write the ODE system (\ref{lcp_ode_2}a) in splitted form as
\begin{equation*}
U'(t) = F(t,U(t)) + G(t,U(t))
\end{equation*}
with
\begin{equation*}
F(t,V) = AV~~ {\rm and}~~ G(t,V)=\lambda(t) 
\quad (0 \le t \le T,~ V\in\R^M).
\end{equation*}
Assume $\hU_{n-1} \approx U(t_{n-1})$ is given and consider $\hU_{n} \approx U(t_{n})$ defined by 
\begin{equation}\label{Do_2}
\begin{cases}
Y_0\, = \hU_{n-1} + \Delta t\, F(t_{n-1},\hU_{n-1}) + \Delta t\, G(t_{n-1},\hU_{n-1}),\phantom{\left( \hU_{n-1}) \right)}\\
Y\phantom{_1} = Y_0 + \theta_1 \Delta t \left(F(t_{n},Y) - F(t_{n-1},\hU_{n-1}) \right),\\
Z~\, = Y\, + \theta_2 \Delta t \left(G(t_{n},Z) - G(t_{n-1},\hU_{n-1}) \right),\\
\hU_{n} = Z.
\end{cases}
\end{equation}
The above can be viewed as a {\it Douglas type splitting}\, scheme for (\ref{lcp_ode_2}a) involving two parameters $\theta_1$,~$\theta_2$, compare~e.g.~\cite{KR_HV03}.
Note that the splitting here is different from the directional splitting considered in Section~\ref{ADIschemes}.
A simple relation holds between the scheme \eqref{Do_2} and the $\theta$-IT method (\ref{LCP_IT}): upon taking $\theta_1=\theta$ and $\theta_2=1$, writing $Y=\bU_n$ and replacing $\lambda(t_q)$ by an approximation $\hl_q$ for $q\in\{n-1,n\}$, it is easily seen that \eqref{Do_2} becomes (\ref{LCP_IT}a) together with the first line of (\ref{LCP_IT}b).
The second line of (\ref{LCP_IT}b), which complements the $\theta$-IT method, forms a discrete analogue of the complementarity condition (\ref{lcp_ode_2}b) at $t=t_n$.
We mention that a related, operator-theoretic derivation was given~in~\cite{KR_LM79} if $\theta=1$, where it was called the Douglas--Rachford scheme. 

The above interpretation of the $\theta$-IT method is directly extended to all \mbox{ADI-IT} methods \eqref{Do}, \eqref{MCS}, \eqref{HV} upon nesting into \eqref{Do_2} the directional splitting of the function $F$ induced by \eqref{splitA}.

Consider next a {\it Peaceman--Rachford type splitting}\, scheme for (\ref{lcp_ode_2}a),
\begin{equation}\label{PR_2}
\begin{cases}
Y\phantom{_1} = \hU_{n-1} + \tfrac{1}{2}\Delta t\, F(t_{n-1/2},Y) +  \tfrac{1}{2} \Delta t\, G(t_{n-1},\hU_{n-1}),\phantom{\left( \hU_{n-1}) \right)}\\
Z~\, = Y~~~~\, + \tfrac{1}{2}\Delta t\, F(t_{n-1/2},Y) + \tfrac{1}{2} \Delta t\, G(t_{n},Z),\phantom{\left( \hU_{n-1}) \right)}\\
\hU_{n} = Z.\phantom{\left( \hU_{n-1}) \right)}
\end{cases}
\end{equation}
Elaborating \eqref{PR_2}, and next replacing $\lambda(t_q)$ by an approximation $\hl_q$ for $q\in\{n-1,n\}$, gives
\begin{equation*}
\hU_{n} = (I+\tfrac{1}{2}\Delta t A)\bU_n + \tfrac{1}{2}\Delta t\, \hl_{n} \quad {\rm with}~ \bU_n {\rm ~defined~by}~ (\ref{LCP_PR}{\rm a}).
\end{equation*}
The discrete analogue of the complementarity condition (\ref{lcp_ode_2}b) at $t=t_n$ reads
\begin{equation*}
\hU_n\geq U_0,~~ \hl_n\geq 0,~~ (\hU_n-U_0)^{\rm T}\, \hl_n =0.
\end{equation*}
This is equivalent, for any given $\varepsilon>0$, to
\begin{equation*}
\hU_n - U_0 = \max \left\{ 0 \,,\, \hU_n - U_0 - \varepsilon \hl_n \right\} ~~{\rm and} ~~
\hl_n = \max \left\{ 0 \,,\, \hl_n - (\hU_n - U_0)/\varepsilon \right\}.
\end{equation*}
Selecting $\varepsilon = \tfrac{1}{2}\Delta t$ and inserting the above expression for $\hU_{n}$  yields (\ref{LCP_PR}b).
Hence, the PR method (\ref{LCP_PR}) can be viewed as obtained from a Peaceman--Rachford type splitting scheme, with the comment that the pertinent splitting is not directional.
This interpretation corresponds to the operator-theoretic exposition given~in~\cite{KR_LM79}. 

We remark that a natural variant to the $\theta$-IT method is obtained by selecting $\theta_1=\theta_2=\theta$ in \eqref{Do_2}.
This leads to (\ref{LCP_IT}) except that in the first line of the update (\ref{LCP_IT}b) the step size $\Delta t$ is replaced by $\theta \Delta t$.
Accordingly, the same replacement occurs in \eqref{LCP_IT_update}.
As it turns out, for $\theta=\tfrac{1}{2}$ this variant of the $\theta$-IT method is equivalent to the PR method.

\section{Numerical study}\label{numerical_study}
In the following we present extensive numerical experiments for the temporal discretization methods described in Section \ref{temporal}.
Our main objectives are to study their actual convergence behaviour in the numerical solution of \eqref{lcp_ode} and to assess their relative performance.

To this purpose, we study the {\it temporal discretization error}\, at $t=T=N \Delta t$, on a natural region of interest, defined by
\begin{equation}\label{temp_error}
\he (\Delta t;m) = \max \{ |U_l(T)-\hU_{N,l}|:\, 0\le i,j\le m,~\tfrac{1}{2}K < s_{1,i}\,,\, s_{2,j} < \tfrac{3}{2}K \}.
\end{equation}
Here $U(T)$ represents the exact solution to the semidiscrete PDCP \eqref{lcp_ode} for $t=T$ and $l=l(i,j)$ denotes the index such that the components $U_l(T)$ and $\hU_{N,l}$ correspond to the spatial grid point $(s_{1,i}\,,\,s_{2,j})$.
In our experiments always the same number of mesh points in the two spatial directions is taken, $m_1=m_2=m$.
For each given $m$, a reference solution for $U(T)$ is computed by applying the $\theta$-P method with $\theta=\tfrac{1}{2}$ and $N=10m$ time steps.

Clearly, \eqref{temp_error} measures the temporal error in the maximum norm, which is the most relevant norm in financial practice.
Note that the spatial discretization error is not contained in \eqref{temp_error}.
We investigate here in detail the error due to the temporal discretization itself. This will lead to important new insights.
We take the number of time steps $N$ directly proportional to $m$, which forms the common situation in applications.
The following methods are considered:\\\\
\noindent
\begin{tabular}{ll}
$\bullet$~~\mbox{BE-EP}\,: & \eqref{LCP_explicit} with $\theta=1$ \\
$\bullet$~~\mbox{BE-IT}\,: & \eqref{LCP_IT} with $\theta=1$ \\
$\bullet$~~\mbox{BE-P}\,:  & \eqref{LCP_penalty} with $\theta=1$ \\
$\bullet$~~\mbox{CN-EP}\,: & \eqref{LCP_explicit} with $\theta=1/2$ \\
$\bullet$~~\mbox{CN-IT}\,: & \eqref{LCP_IT} with $\theta=1/2$ \\
$\bullet$~~\mbox{CN-P}\,:  & \eqref{LCP_penalty} with $\theta=1/2$ \\
$\bullet$~~PR\,:    & \eqref{LCP_PR}
\end{tabular}\\\\
and\\\\
\noindent
\begin{tabular}{ll}
$\bullet$~~\mbox{Do-IT}\,: & \eqref{Do} with $\theta=1/2$ \\
$\bullet$~~\mbox{CS-IT}\,:& \eqref{MCS} with $\theta=1/2$ \\
$\bullet$~~\mbox{MCS-IT}\,:& \eqref{MCS} with $\theta=1/3$ \\
$\bullet$~~\mbox{HV-IT}\,: & \eqref{HV} with $\theta=1/(2+\sqrt{2})$
\end{tabular}\\\\
The selected values of $\theta$ for methods \eqref{Do}, \eqref{MCS}, \eqref{HV} are motivated by the favourable unconditional stability results obtained for the underlying ADI schemes in \cite{KR_HW07,KR_HW09}.

It is well-known that in financial applications the payoff function $\phi$ is usually nonsmooth at one or more given points, which has an adverse effect on the accuracy of numerical solution methods.
For the spatial discretization, this effect can be alleviated by constructing a spatial grid such that the coordinates of these points of nonsmoothness always lie exactly midway between two successive mesh points.
Such a construction has been considered in Section~\ref{spatial_discr}.
Subsequently, for the temporal discretization, a common approach is to apply backward Euler damping, also known as Rannacher time stepping. 
In the case of European options, this means that the first few time steps are all replaced by two substeps with step size $\Delta t/2$ of the backward Euler method.
In analogy to this, we always replace each of the first two time steps of any of the $\theta$-EP, $\theta$-IT and $\theta$-P methods by two substeps with step size $\Delta t/2$ of the same method using $\theta=1$.
Next, for damping the PR method and all \mbox{ADI-IT} methods, the $\theta$-IT method is employed with $\theta=1$.

\subsection{One-asset American options}\label{oneasset}
We begin with the special case of one-asset American options under the Black--Scholes framework.
The pertinent (one-dimensional) spatial differential operator is 
\begin{equation}\label{calA_2}
\mathcal{A}   =
\tfrac{1}{2} \sigma^2 s^2 \frac{\partial^2 }{\partial s^2}
+ r s \frac{\partial }{\partial s}
- r.
\end{equation}
The spatial discretization is performed as in Section~\ref{spatial_discr}, where for the nonuniform mesh the following parameter values are taken,
\begin{equation}\label{gridpars}
d=K/3,~~S_{\lleft}=0.8K,~~S_{\rright}=1.2K,~~S_{\rm max}=5K.
\end{equation}

\begin{figure}[!]
\begin{center}
 \includegraphics[width=0.7\textwidth]{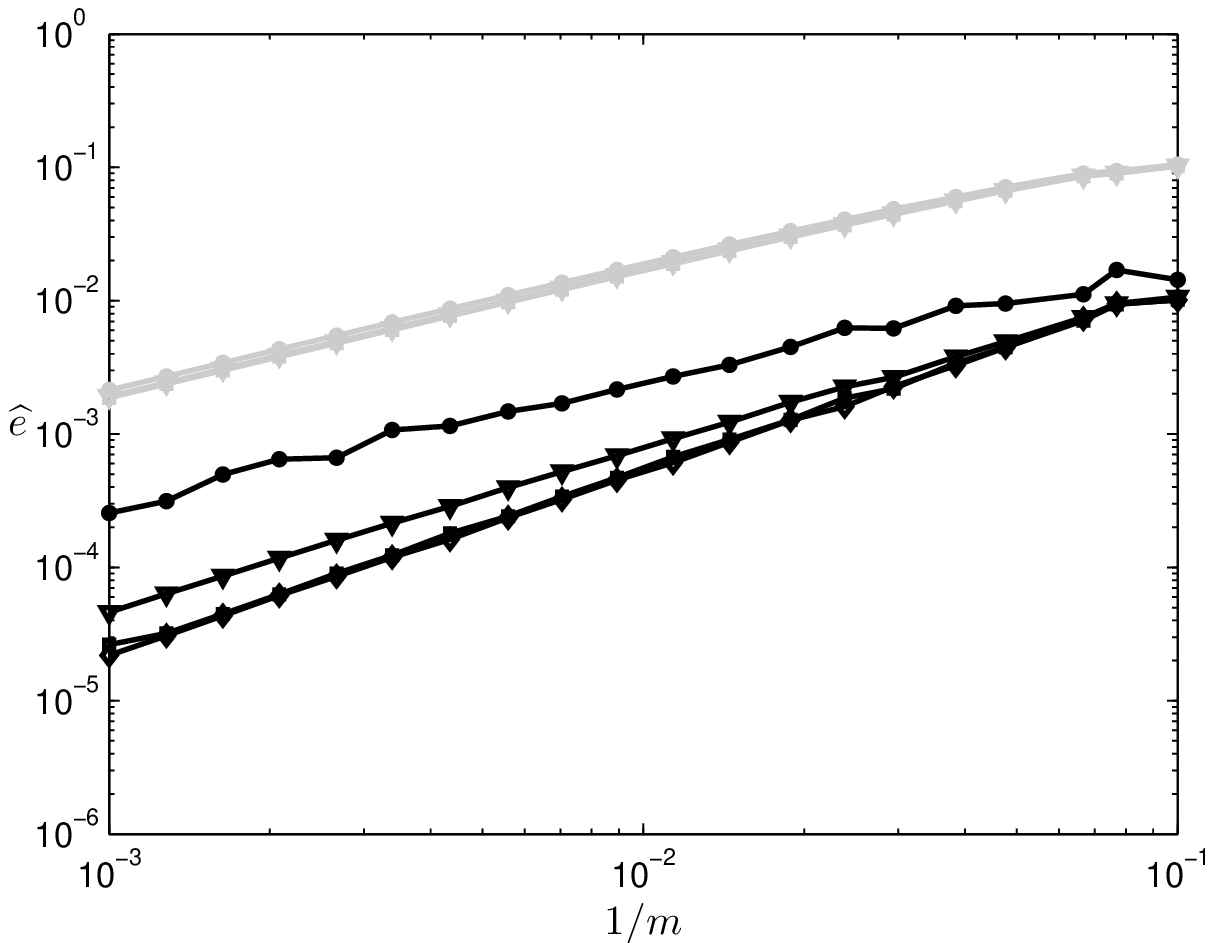}
\end{center}
\caption{American put option and parameters \eqref{1Dputpar}. Temporal error $\he(\Delta t;m)$ versus $1/m$ with $N=m$ for $10\le m \le 1000$. Constant step sizes. \mbox{BE-EP}: light bullets, \mbox{BE-IT}: light squares, \mbox{BE-P}: light triangles, \mbox{CN-EP}: dark bullets, \mbox{CN-IT}: dark squares, \mbox{CN-P}: dark triangles, PR: dark diamonds.}
\label{Fig1Dputcon}

\begin{center}
 \includegraphics[width=0.7\textwidth]{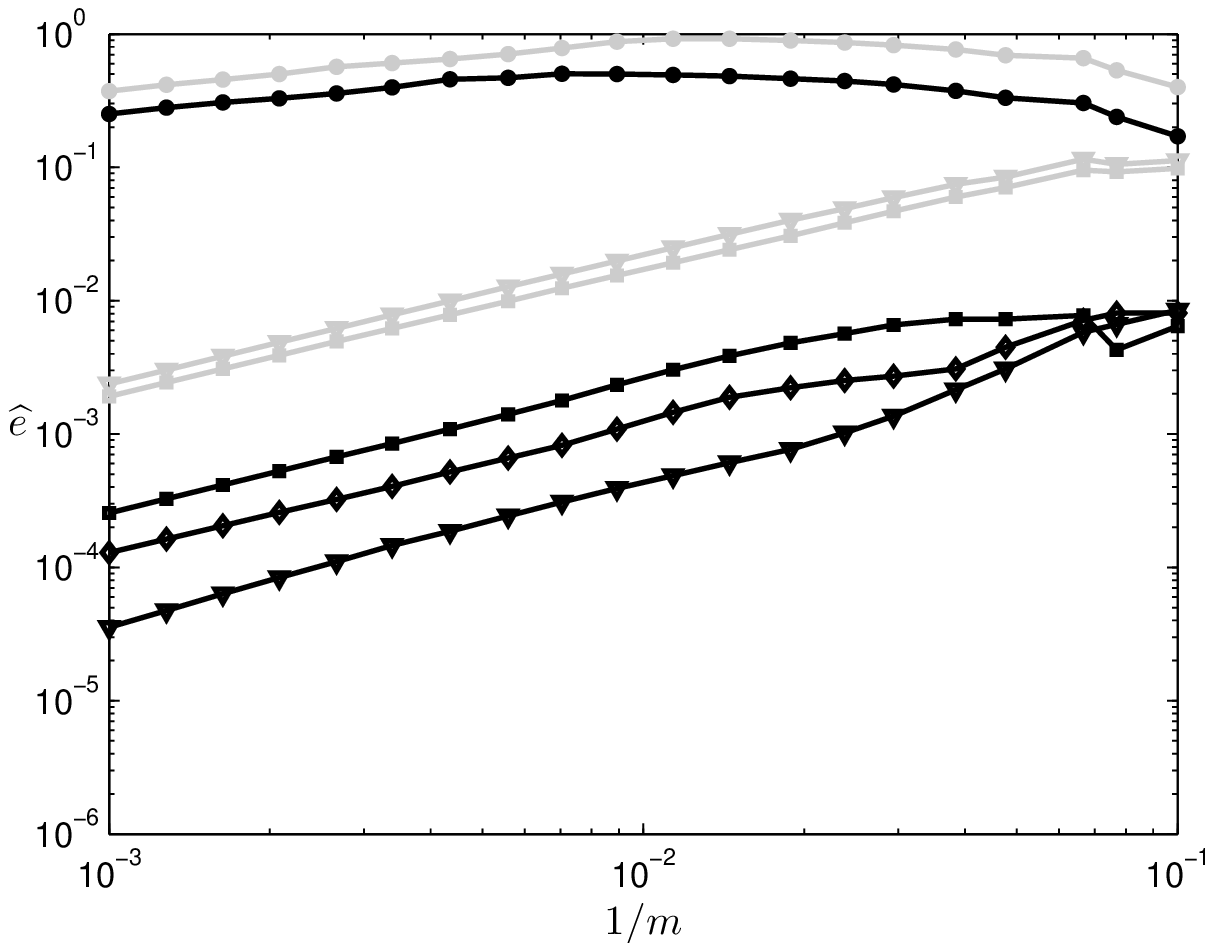}
\end{center}
\caption{American butterfly option and parameters \eqref{1Dbutpar}. Temporal error $\he(\Delta t;m)$ versus $1/m$ with $N=m$ for $10\le m \le 1000$. Constant step sizes. \mbox{BE-EP}: light bullets, \mbox{BE-IT}: light squares, \mbox{BE-P}: light triangles, \mbox{CN-EP}: dark bullets, \mbox{CN-IT}: dark squares, \mbox{CN-P}: dark triangles, PR: dark diamonds.}
\label{Fig1Dbutcon}
\end{figure}

As a first example we consider an American put option, which has payoff $\phi(s) = \max (K-s,0)$ (for $s\ge 0$), and choose financial parameter values
\begin{equation}\label{1Dputpar}
r=0.02,~~\sigma = 0.40,~~T=0.5,~~K=100.
\end{equation}
Figure~\ref{Fig1Dputcon} displays, for all methods listed above except the \mbox{ADI-IT} methods, their temporal discretization errors $\he (\Delta t;m)$ for $N=m$ and 20 different values $m$ between 10 and 1000, given by an equal number of appropriate odd values $\nu$ (see Section~\ref{spatial_discr}).
 One observes that the errors obtained with the three methods \mbox{BE-EP}, \mbox{BE-IT}, \mbox{BE-P} are very close to each other.
They show a first-order convergence behaviour, as might be expected.
The errors obtained with the four methods \mbox{CN-EP}, \mbox{CN-IT}, \mbox{CN-P}, PR are substantially smaller.
Of these four, the \mbox{CN-EP} method is the least accurate.
The errors for the \mbox{CN-IT}, \mbox{CN-P}, PR methods are relatively close to each other and a convergence order approximately equal to 1.3 is observed for these.
Clearly, this order is significantly lower than two, which is attributed to the nonsmoothness of the option value function near the early exercise boundary, see e.g.~\cite{KR_FV02}.

We next choose the more challenging example of an American butterfly option, see \cite{KR_HRW13}.
It has the nonconvex payoff 
\[
\phi(s) = \max(s-K_1,0)-2\max(s-K,0)+\max(s-K_2,0)
\]
with strikes $K_1<K_2$ and $K=(K_1+K_2)/2$. 
Figure~\ref{Fig1Dbutcon} displays, analogously to the above, the temporal discretization errors for all methods under consideration with financial parameter values
\begin{equation}\label{1Dbutpar}
r=0.02,~~\sigma = 0.40,~~T=0.5,~~K_1=80,~~K_2=120.
\end{equation}
The \mbox{BE-IT}, \mbox{BE-P}, \mbox{CN-IT}, \mbox{CN-P}, PR methods reveal a neat first-order convergence behaviour.
The explicit payoff methods, \mbox{BE-EP} and \mbox{CN-EP}, invariably yield large errors and appear to converge only very slowly as $N=m$ increases.
We mention that for the latter methods the temporal errors are largest near the strike $K$ (which is always an early exercise point for the butterfly option).

Ample additional experiments in the case of one-asset American put and butterfly options support our above conclusions.
The CN-based methods are in general more accurate than the BE-based methods, as could be expected.
Further, it is found that the PR method is often somewhat more accurate than the \mbox{CN-IT} method.

\begin{figure}[!]
\begin{center}
 \includegraphics[width=0.8\textwidth]{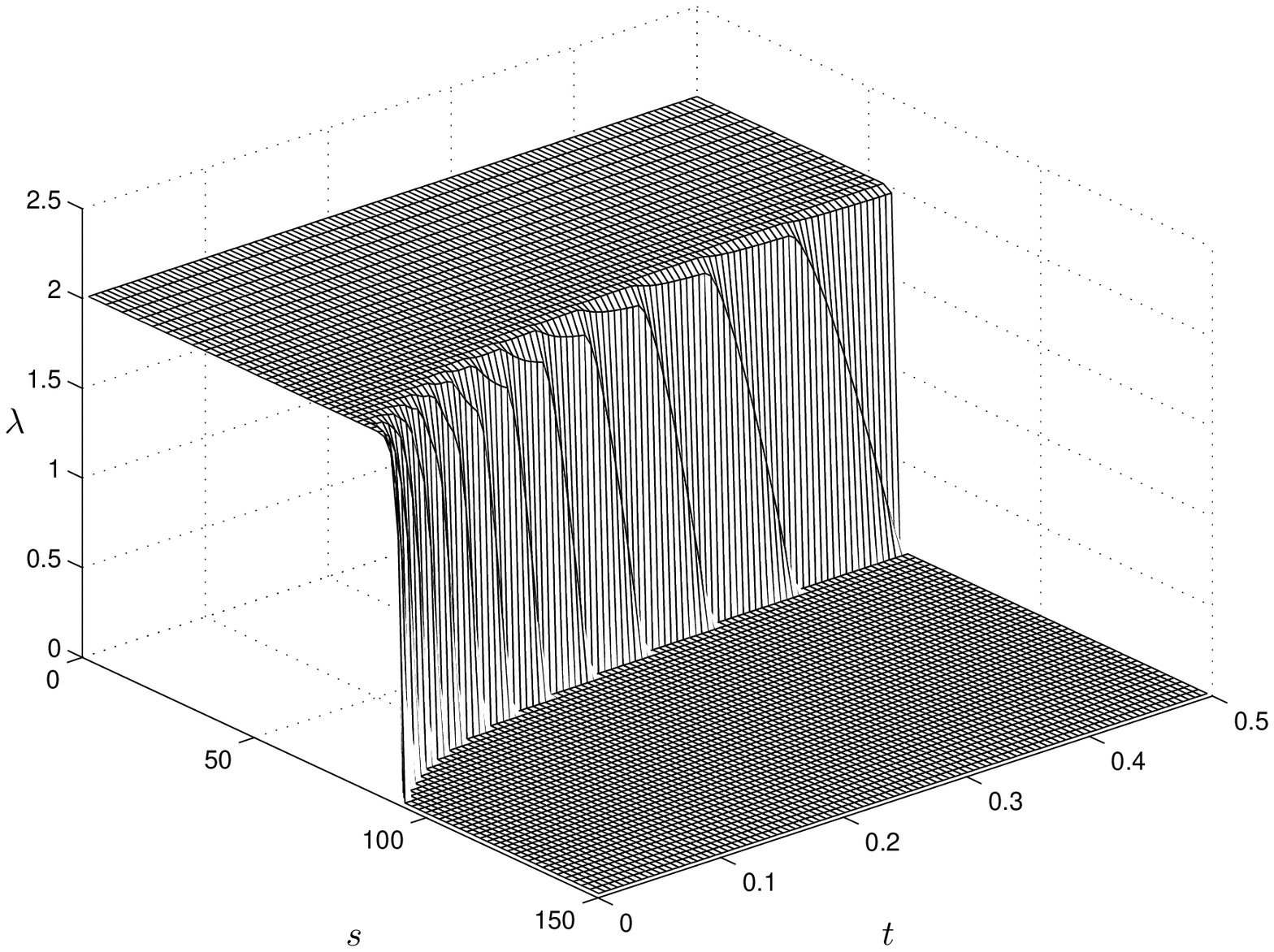}
\end{center}
\caption{American put option and parameters \eqref{1Dputpar}. Lagrange multipliers $\hl_{n,i}$ versus $(s_{1,i},t_n)$ in $[0,\frac{3}{2}K]\times(0,T]$ for \mbox{BE-IT} method and $m=100$.}
\label{Fig1DputLM}

\begin{center}
 \includegraphics[width=0.8\textwidth]{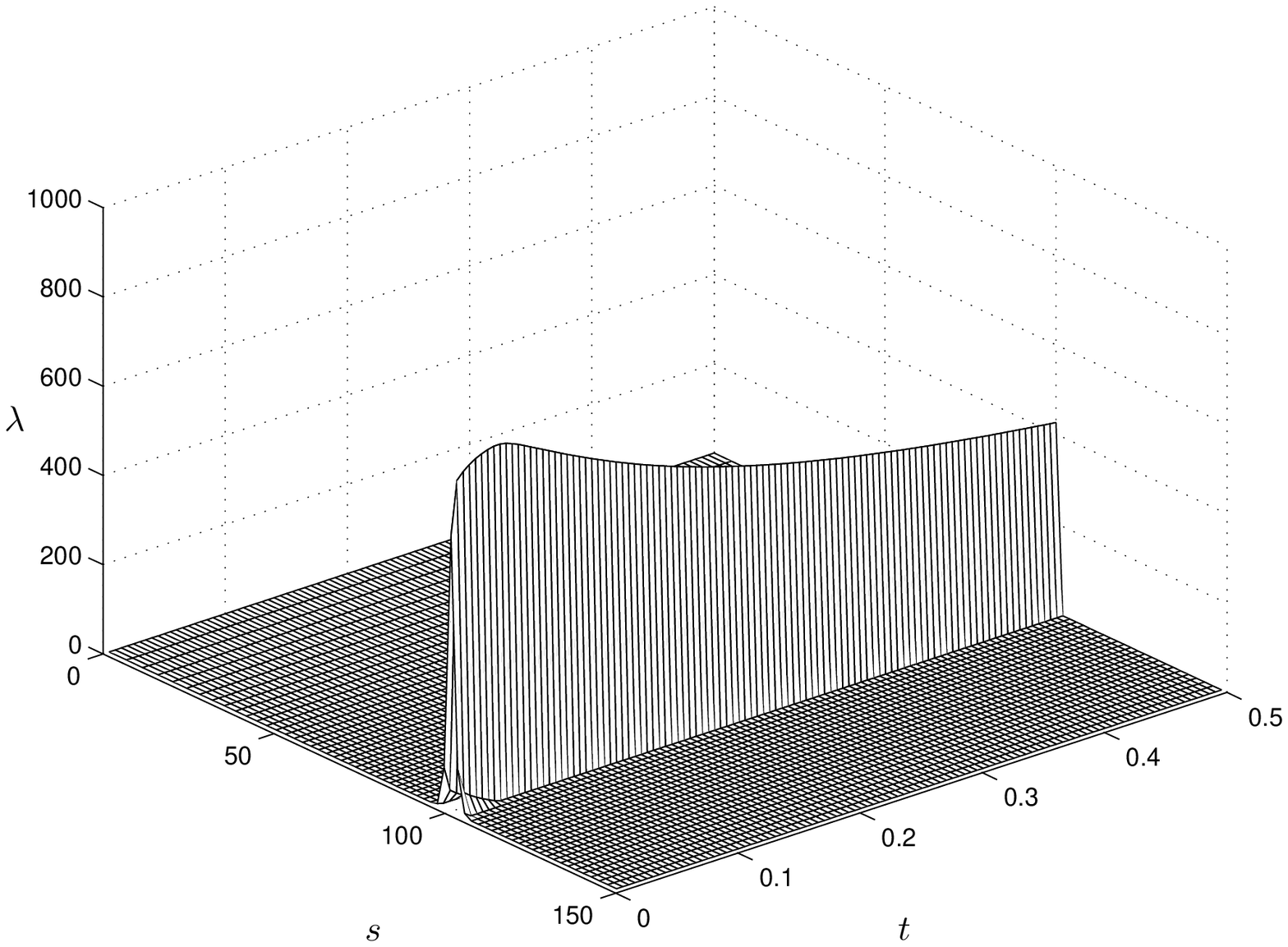}
\end{center}
\caption{American butterfly option and parameters \eqref{1Dbutpar}. Lagrange multipliers $\hl_{n,i}$ versus $(s_{1,i},t_n)$ in $[0,\frac{3}{2}K]\times(0,T]$ for \mbox{BE-IT} method and $m=100$.}
\label{Fig1DbutLM}
\end{figure}

\begin{figure}[!]
\begin{center}
 \includegraphics[width=0.7\textwidth]{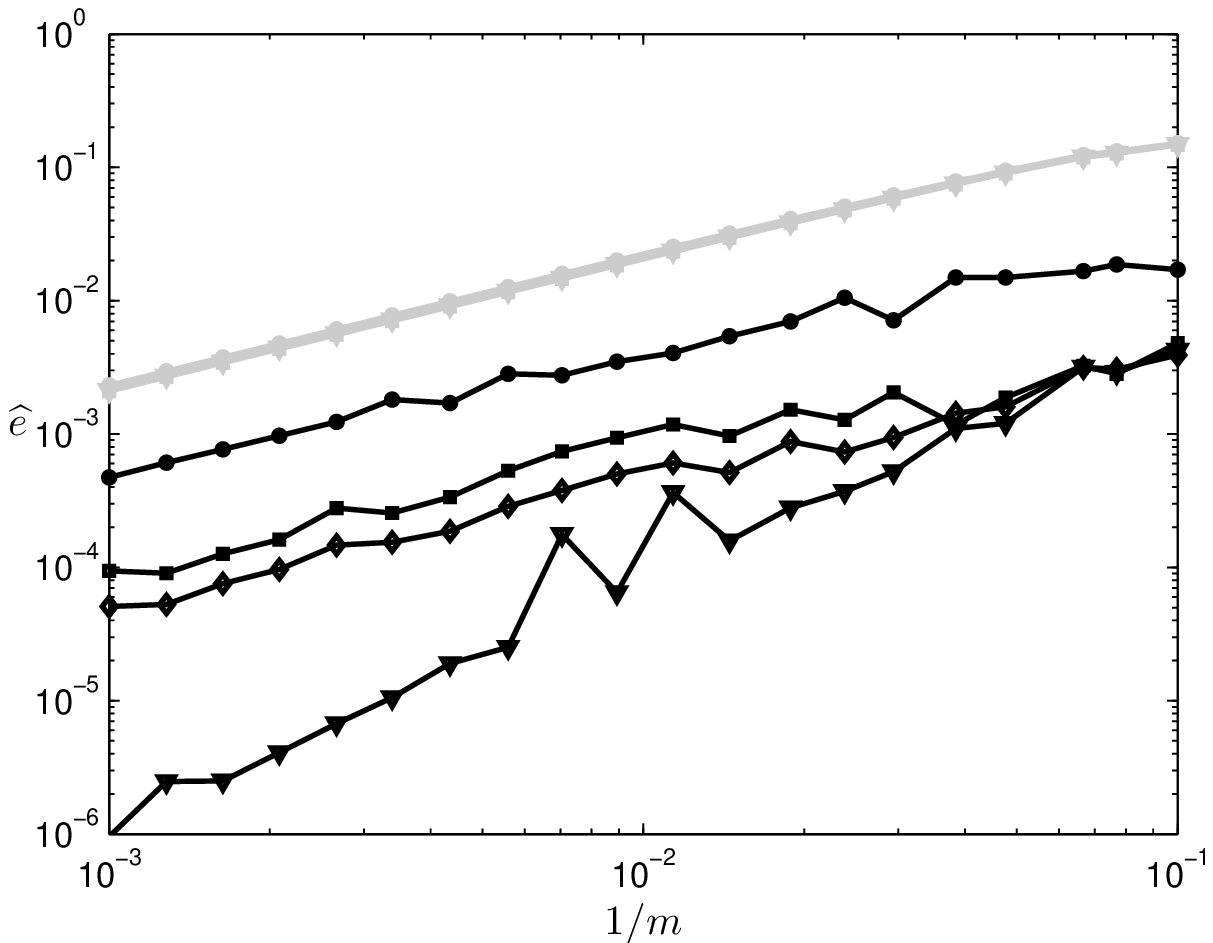}
\end{center}
\caption{American put option and parameters \eqref{1Dputpar}. Temporal error $\he(\Delta t;m)$ versus $1/m$ with $N=m$ for $10\le m \le 1000$. Variable step sizes. \mbox{BE-EP}: light bullets, \mbox{BE-IT}: light squares, \mbox{BE-P}: light triangles, \mbox{CN-EP}: dark bullets, \mbox{CN-IT}: dark squares, \mbox{CN-P}: dark triangles, PR: dark diamonds.}
\label{Fig1Dputvar}

\begin{center}
 \includegraphics[width=0.7\textwidth]{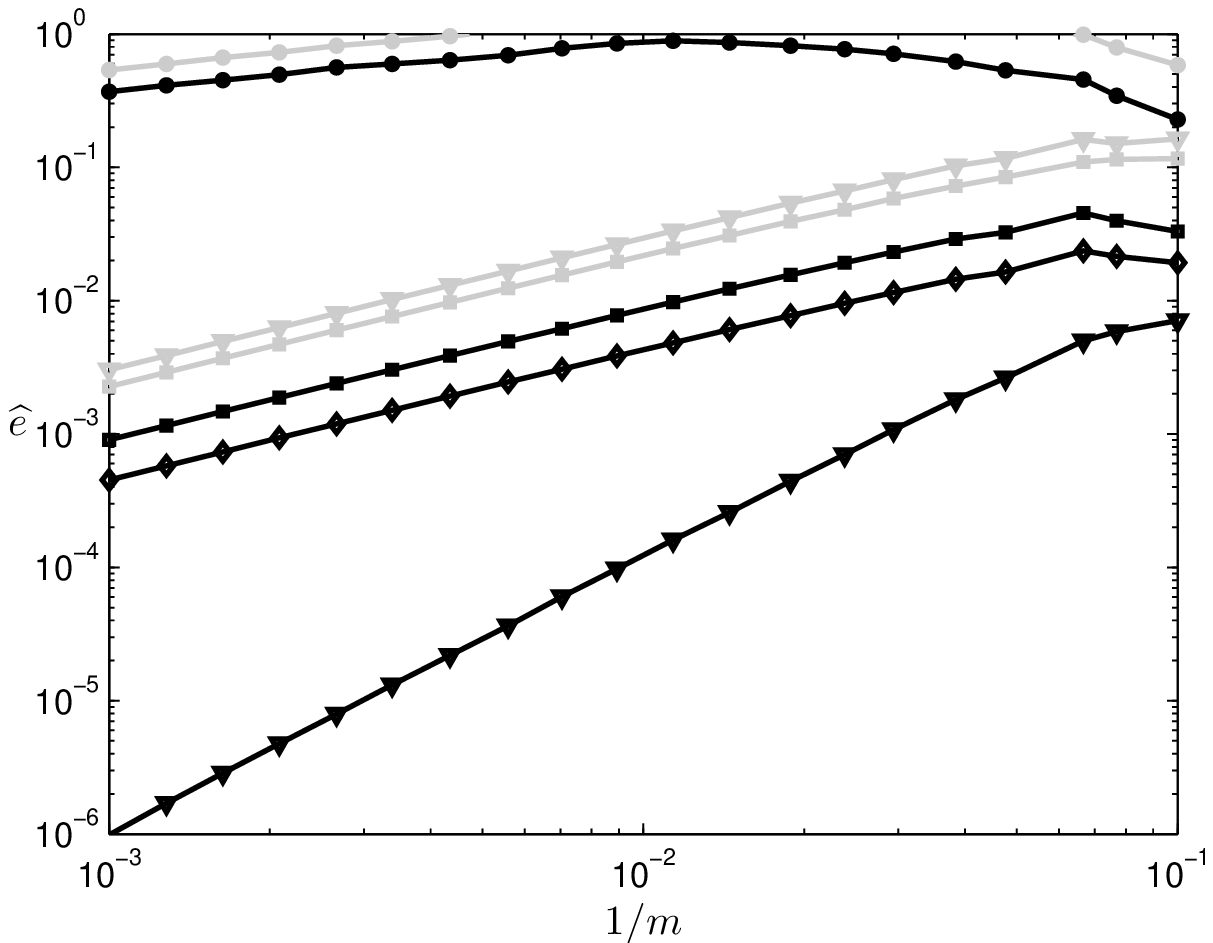}
\end{center}
\caption{American butterfly option and parameters \eqref{1Dbutpar}. Temporal error $\he(\Delta t;m)$ versus $1/m$ with $N=m$ for $10\le m \le 1000$. Variable step sizes. \mbox{BE-EP}: light bullets, \mbox{BE-IT}: light squares, \mbox{BE-P}: light triangles, \mbox{CN-EP}: dark bullets, \mbox{CN-IT}: dark squares, \mbox{CN-P}: dark triangles, PR: dark diamonds.}
\label{Fig1Dbutvar}
\end{figure}

Examining the Lagrange multiplier vectors $\hl_n$ provides further useful insight.
Figures~\ref{Fig1DputLM} and~\ref{Fig1DbutLM} display for the American put and butterfly options, respectively, the Lagrange multiplier $\hl_{n,i}$ versus $(s_{1,i},t_n)$ in the domain $[0,\frac{3}{2}K]\times(0,T]$ for the \mbox{BE-IT} method and $m=100$.
The subdomain where the multiplier is nonzero represents the early exercise region.
Clearly, for the butterfly option this region forms a narrow neighbourhood of the strike $K$.
Replacing the \mbox{BE-IT} method by the \mbox{CN-IT} method or the PR method yields essentially the same outcome as in Figures~\ref{Fig1DputLM} and~\ref{Fig1DbutLM}.
Upon increasing $m$, the outcome for the American put remains approximately the same, but for the American butterfly the maximum grows, in a manner directly proportional to $m$.
The latter phenomenon can be explained from the nonsmoothness (kink) of the exact butterfly option value function at the strike $K$ at all times, which renders the numerical valuation of the American butterfly much more challenging than that of the American put.

It was demonstrated in \cite{KR_FV02} that by employing suitable adaptive variable step sizes, instead of constant step sizes, one can recover second-order convergence for the \mbox{CN-P} method.
All temporal discretization methods from Section~\ref{temporal} are extended straightforwardly to variable step sizes.
We consider here temporal grid points defined upfront by (compare also e.g.~\cite{KR_IT09,KR_RW14})
\begin{equation}\label{variablestep}
t_n = \left( \frac{n}{N} \right)^2 T \quad {\rm for}~ n=0,1,2,\ldots,N.
\end{equation}
The corresponding step sizes are smallest near $t=0$ (which is where the option value and early exercise boundary vary strongest) and they grow linearly with~$n$.
Figures~\ref{Fig1Dputvar} and \ref{Fig1Dbutvar} are the analogues of Figures~\ref{Fig1Dputcon} and \ref{Fig1Dbutcon}, respectively, obtained in the case of these variable step sizes. 
Indeed, for the \mbox{CN-P} method a favourable second-order convergence behaviour is observed.
We note that relatively larger temporal errors can occur near the early exercise boundary in the case of the put option, resulting in the occasional ``peaks'' in Figure~\ref{Fig1Dputvar}, see also \cite{KR_FV02}.
With variable step sizes, however, the \mbox{CN-P} method is in general substantially more accurate than all other methods under consideration.
For the other methods, employing variable step sizes does not lead to a significant improvement in accuracy compared to constant step sizes.
Because with the \mbox{CN-P} method the pertinent LCP in each time step is essentially solved exactly, we conclude that for the other CN-based methods (including PR) the error due to the approximate solution of the LCP in each time step dominates the error due to the CN time stepping; notice that the temporal discretization error \eqref{temp_error} can be viewed as the sum of these two errors, since $U(T)-\hU_{N}=(U(T)-U_N)+(U_N-\hU_{N})$ with $U_N$ defined by \eqref{fullPDCP}.

\subsection{Two-asset American options}
We next consider numerical experiments for several two-asset American options.
The spatial discretization of the PDCP (\ref{PDCP}) is done on the nonuniform grid described in Section~\ref{spatial_discr} with parameter values \eqref{gridpars}.
For the temporal discretization we apply all methods listed at the beginning of this section except those using the explicit payoff  approach. 

\begin{figure}[!]
\begin{center}
 \includegraphics[width=0.6\textwidth]{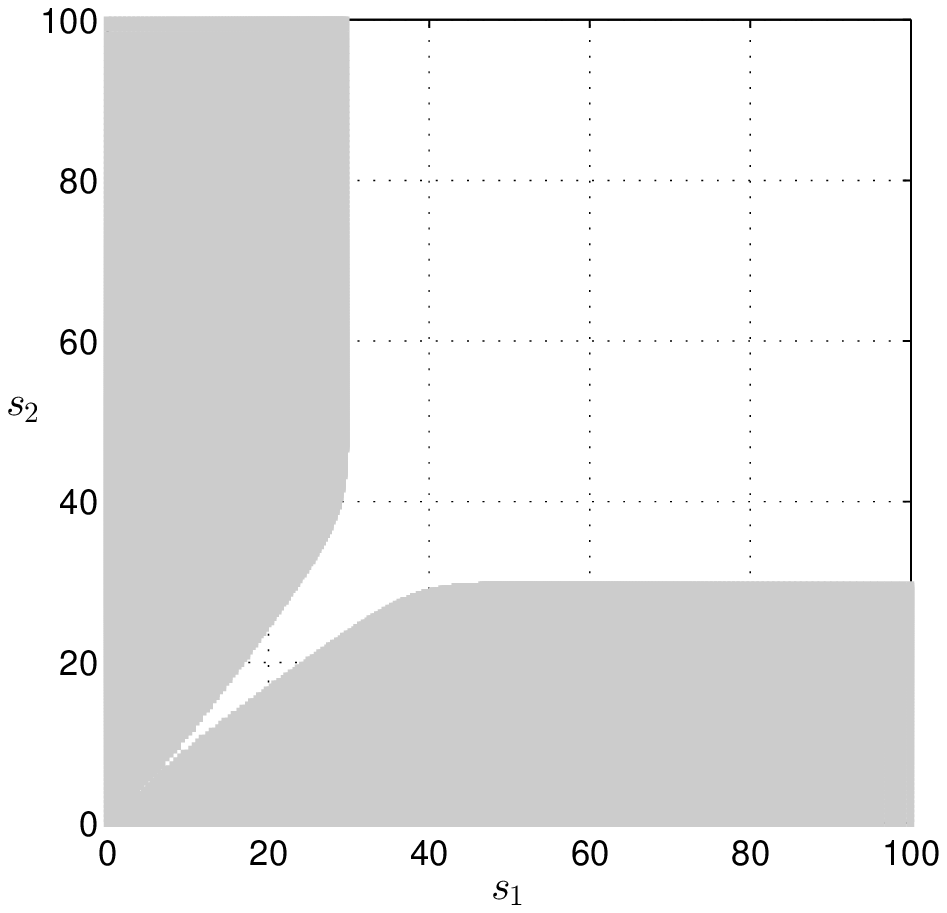}
\end{center}
\caption{Early exercise region if $t=T$ for two-asset American put on the minimum and parameters \eqref{2Dputpar}.}
\label{Fig2DputFB}
\begin{center}
 \includegraphics[width=0.6\textwidth]{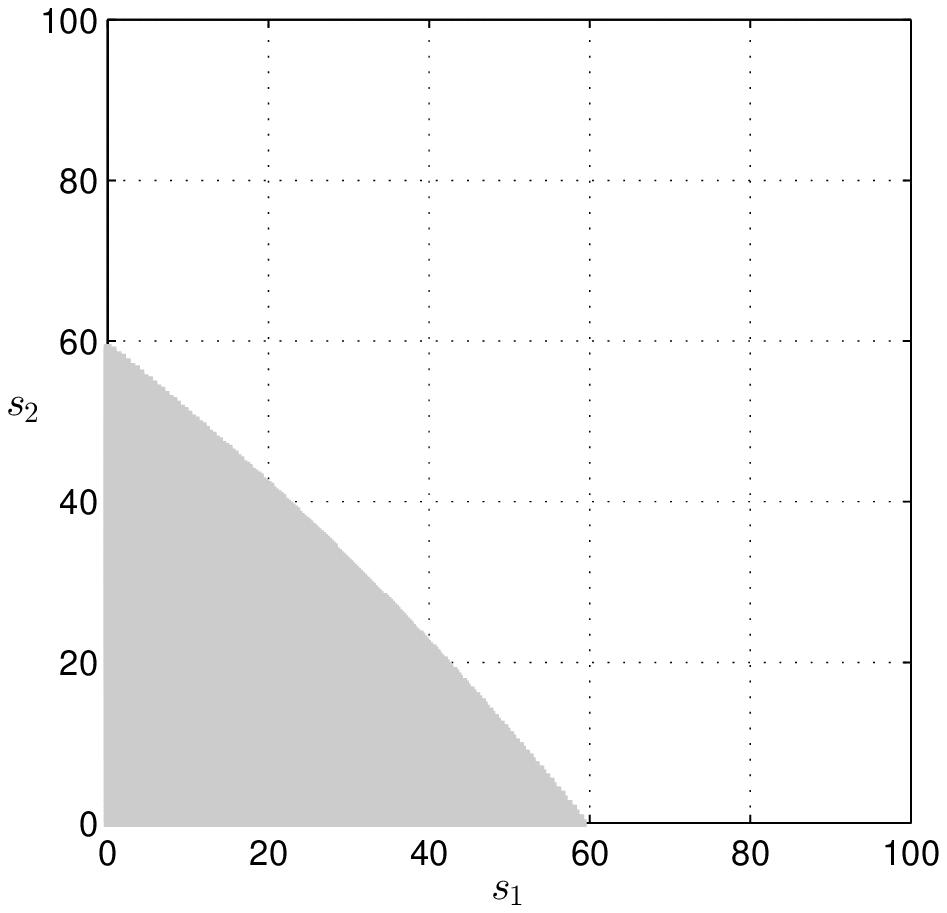}
\end{center}
\caption{Early exercise region if $t=T$ for two-asset American put on the average and parameters \eqref{2Dputpar}.}
\label{Fig2DbasFB}
\end{figure}

\begin{figure}[!]
\begin{center}
 \includegraphics[width=0.7\textwidth]{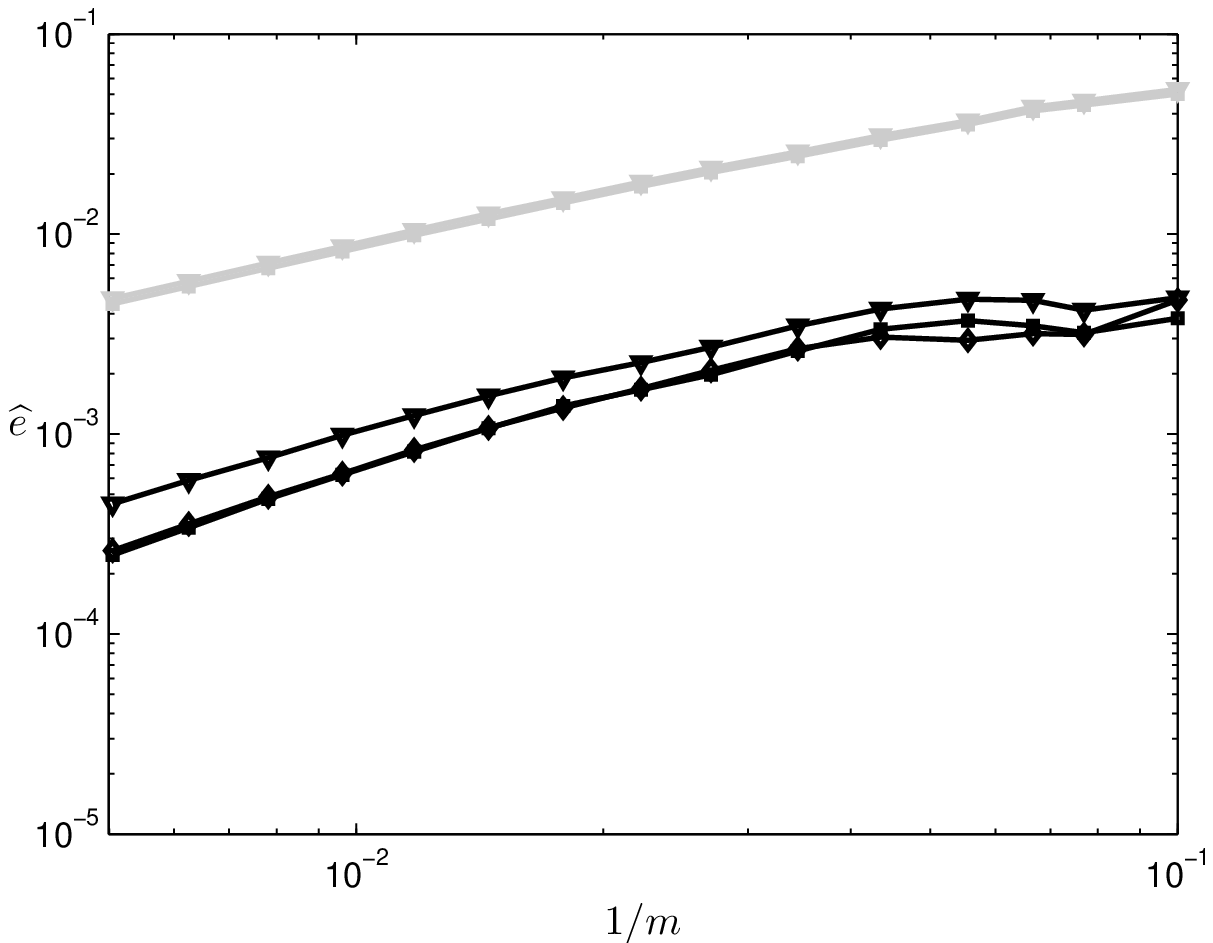}
\end{center}
\caption{Two-asset American put on the minimum and parameters \eqref{2Dputpar}. Temporal error $\he(\Delta t;m)$ versus $1/m$ with $N=m$ for $10\le m \le 200$. Constant step sizes. \mbox{BE-IT}: light squares, \mbox{BE-P}: light triangles, \mbox{CN-IT}: dark squares, \mbox{CN-P}: dark triangles, PR: dark diamonds. }
\label{Fig2Dputcon_theta}
\begin{center}
 \includegraphics[width=0.7\textwidth]{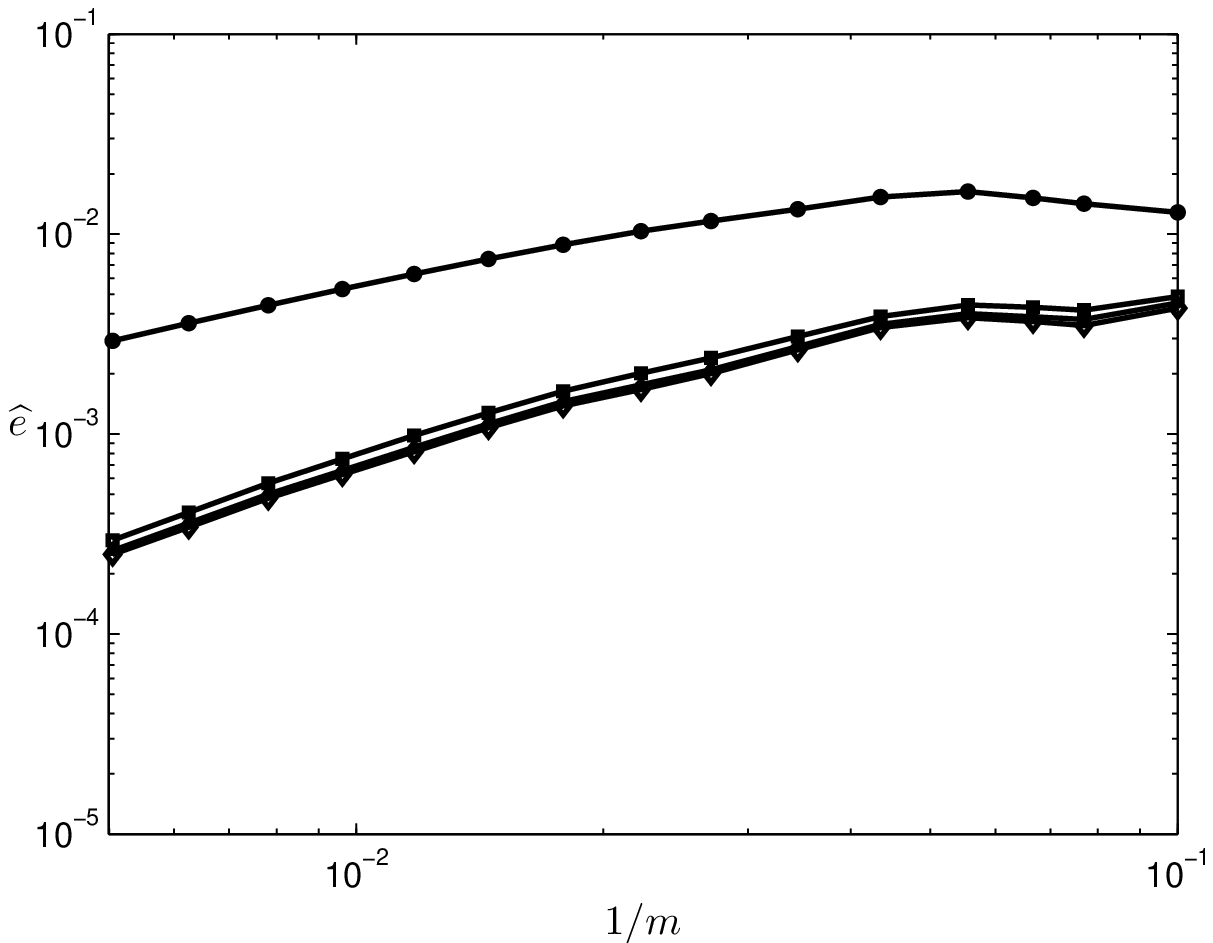}
\end{center}
\caption{Two-asset American put on the minimum and parameters \eqref{2Dputpar}. Temporal error $\he(\Delta t;m)$ versus $1/m$ with $N=m$ for $10\le m \le 200$. Constant step sizes. \mbox{Do-IT}: dark bullets, \mbox{CS-IT}: dark squares, \mbox{MCS-IT}: dark stars, \mbox{HV-IT}: dark diamonds.}
\label{Fig2Dputcon_ADI}
\end{figure}

As a first example an American put option on the minimum of two asset prices is taken.
It has payoff $\phi(s_1,s_2) = \max (K-s,0)$ with $s=\min(s_1,s_2)$.
We choose financial parameter values from \cite{KR_ZFV01},
\begin{equation}\label{2Dputpar}
r=0.05,~~\sigma_1 = 0.30,~~\sigma_2 = 0.30,~~\rho = 0.50,~~T=0.5,~~K=40.
\end{equation}
The numerically approximated early exercise region for $t=T$ is shown in Figure~\ref{Fig2DputFB}.
We compute the temporal discretization errors $\he (\Delta t;m)$ for $N=m$ constant step sizes and 15 different values $m$ between 10 and 200, corresponding to an equal number of odd values $\nu$.
Figure~\ref{Fig2Dputcon_theta} displays the obtained results for the $\theta$-based methods.
Similar to the one-asset American put option case, the errors for the \mbox{BE-IT} and \mbox{BE-P} methods are very close to each other and show an approximate first-order convergence behaviour.
Also, as before, the errors for the CN-based methods are substantially smaller and relatively close to each other and show a convergence order approximately equal to 1.3.
In Figure~\ref{Fig2Dputcon_ADI} the results are displayed for the four \mbox{ADI-IT} methods under consideration.
The obtained accuracies with the \mbox{CS-IT}, \mbox{MCS-IT} and \mbox{HV-IT} methods are about the same and close to those for the CN-based methods.
The observed convergence orders for these three \mbox{ADI-IT} methods are thus also approximately equal to 1.3.
The Do-IT method is substantially less accurate than the three more advanced \mbox{ADI-IT} methods, but it is somewhat more accurate than the \mbox{BE-IT} and \mbox{BE-P} methods. 
The observed convergence order for the Do-IT method is slightly smaller than~one.

\begin{figure}[!]
\begin{center}
 \includegraphics[width=0.7\textwidth]{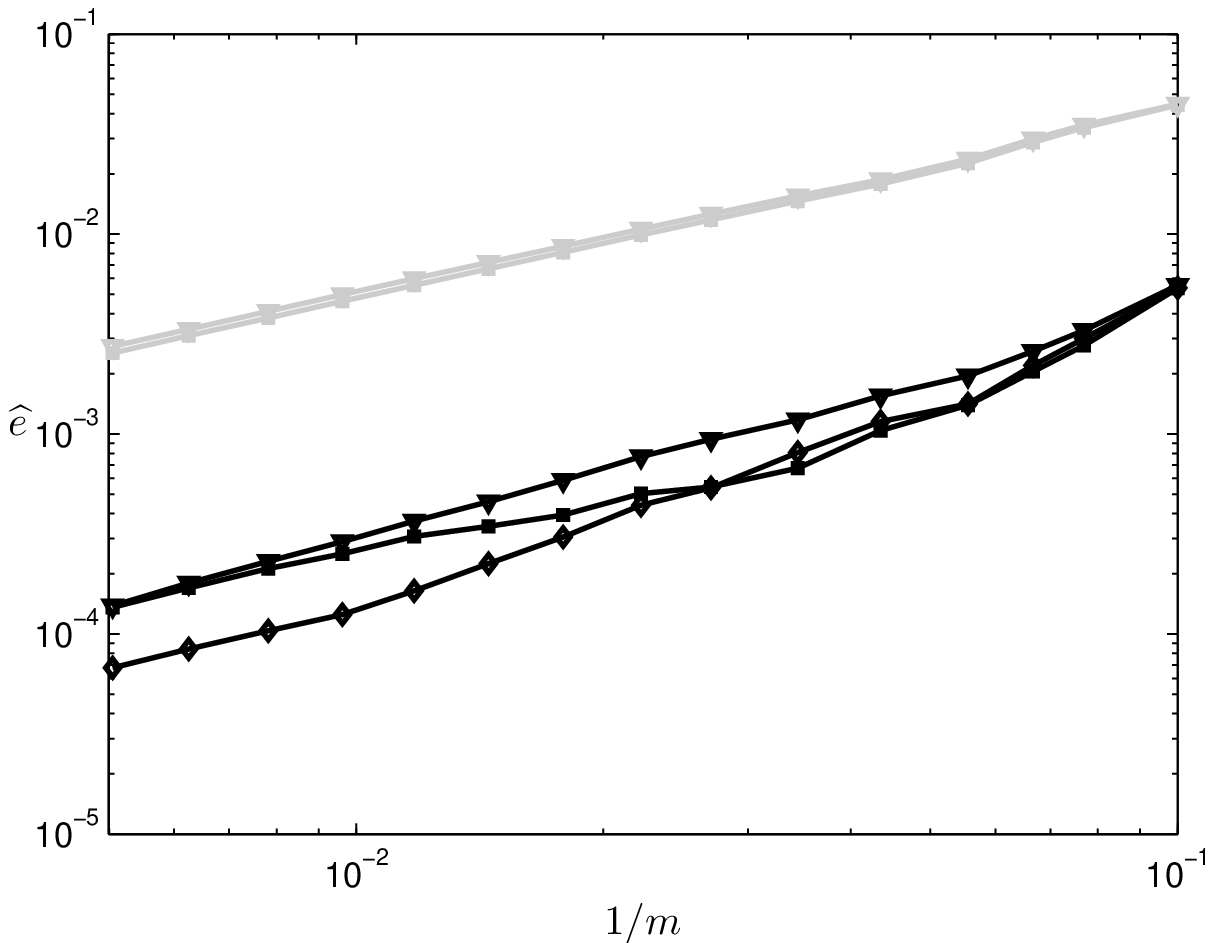}
\end{center}
\caption{Two-asset American put on the average and parameters \eqref{2Dputpar}. Temporal error $\he(\Delta t;m)$ versus $1/m$ with $N=m$ for $10\le m \le 200$. Constant step sizes. \mbox{BE-IT}: light squares, \mbox{BE-P}: light triangles, \mbox{CN-IT}: dark squares, \mbox{CN-P}: dark triangles, PR: dark diamonds.}
\label{Fig2Dbascon_theta}
\begin{center}
 \includegraphics[width=0.7\textwidth]{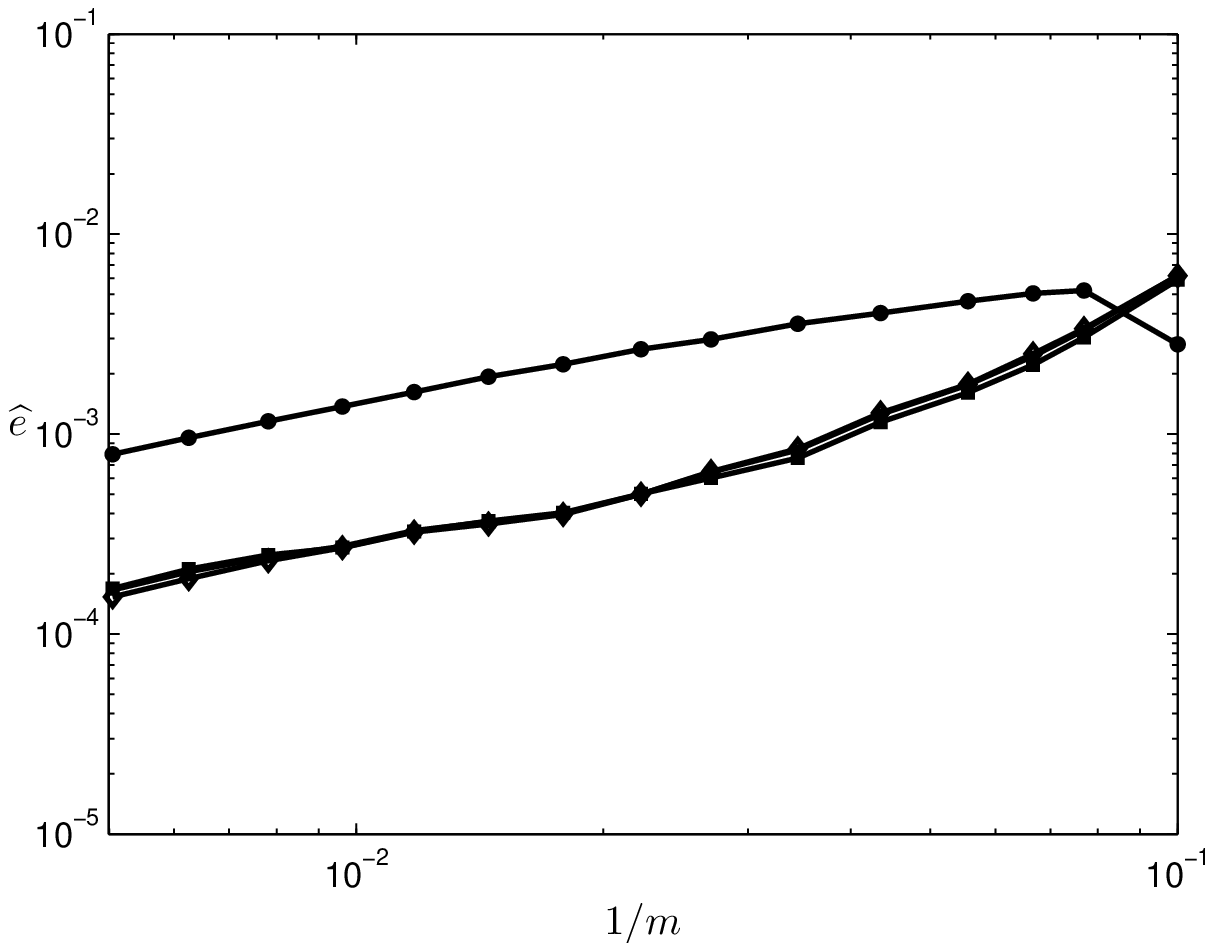}
\end{center}
\caption{Two-asset American put on the average and parameters \eqref{2Dputpar}. Temporal error $\he(\Delta t;m)$ versus $1/m$ with $N=m$ for $10\le m \le 200$. Constant step sizes. \mbox{Do-IT}: dark bullets, \mbox{CS-IT}: dark squares, \mbox{MCS-IT}: dark stars, \mbox{HV-IT}: dark diamonds.}
\label{Fig2Dbascon_ADI}
\end{figure}

As a second example we consider an American put option on the arithmetic average of two asset prices, which has the payoff $\phi(s_1,s_2) = \max (K-s,0)$ with $s=(s_1+s_2)/2$.
The numerically approximated early exercise region when $t=T$ is displayed in Figure~\ref{Fig2DbasFB}. 
Figures~\ref{Fig2Dbascon_theta} and \ref{Fig2Dbascon_ADI} form the analogues of Figures~\ref{Fig2Dputcon_theta} and \ref{Fig2Dputcon_ADI}, respectively, for this option.
Comparing the achieved accuracies of the different methods, the same conclusions are obtained as in the case of the put option on the minimum, with the exception of the PR method, which is often somewhat more accurate than the other methods.
The observed convergence orders for the CN-P and PR methods are approximately equal to 1.1 and 1.3, respectively, and for all other methods they are slightly smaller than one.

\begin{figure}[!]
\begin{center}
 \includegraphics[width=0.7\textwidth]{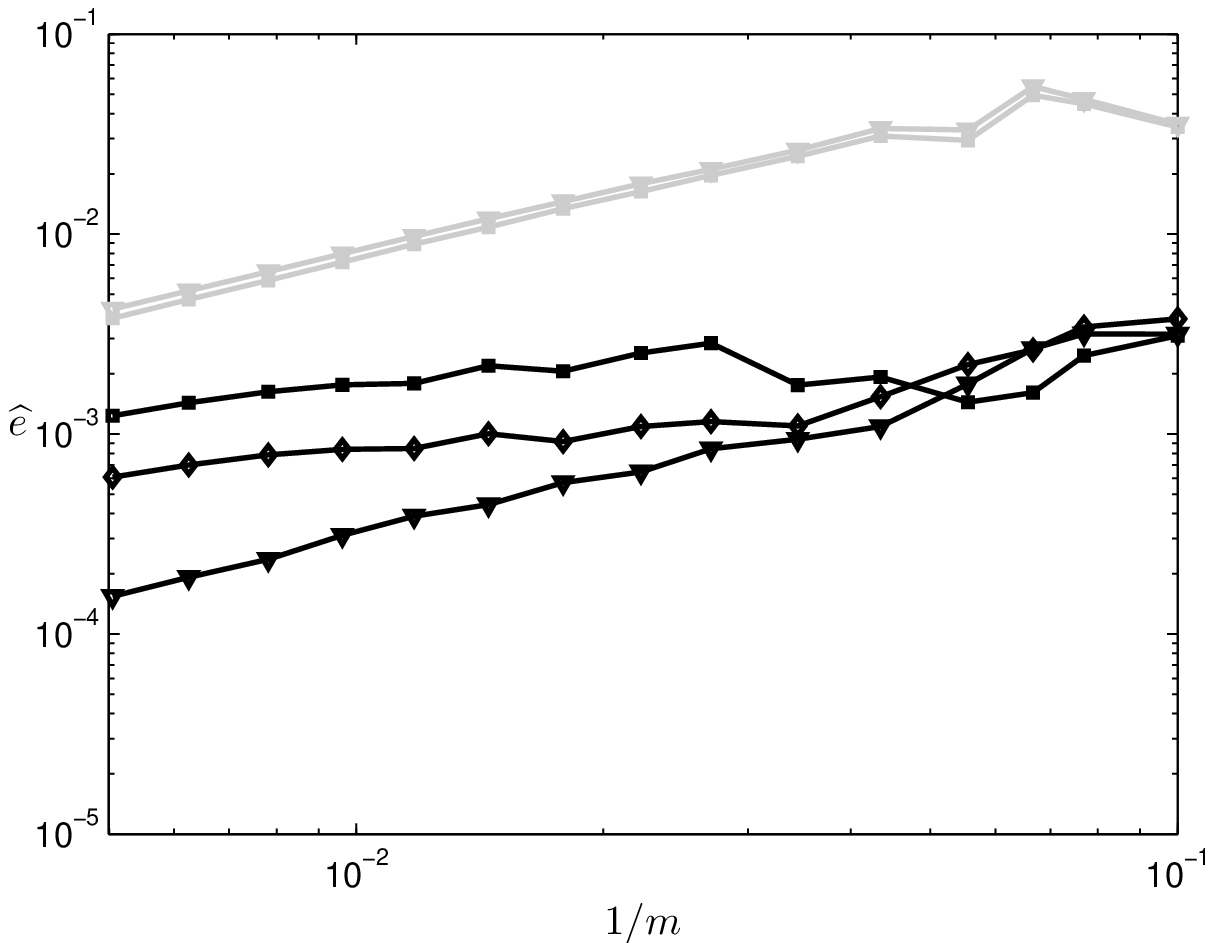}
\end{center}
\caption{Two-asset American butterfly and parameters \eqref{2Dbutpar}. Temporal error $\he(\Delta t;m)$ versus $1/m$ with $N=m$ for $10\le m \le 200$. Constant step sizes. \mbox{BE-IT}: light squares, \mbox{BE-P}: light triangles, \mbox{CN-IT}: dark squares, \mbox{CN-P}: dark triangles, PR: dark diamonds.}
\label{Fig2Dbutcon_theta}
\begin{center}
 \includegraphics[width=0.7\textwidth]{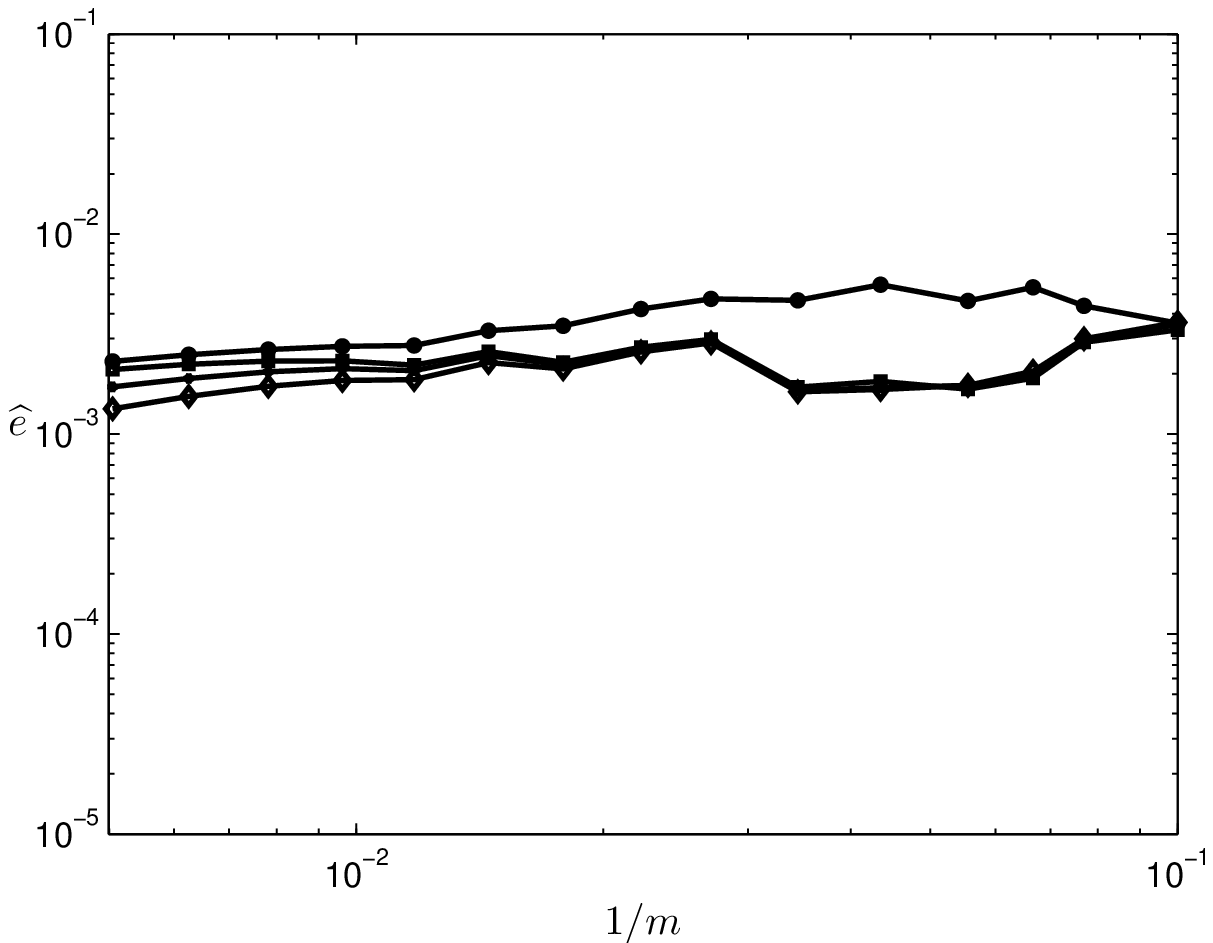}
\end{center}
\caption{Two-asset American butterfly and parameters \eqref{2Dbutpar}. Temporal error $\he(\Delta t;m)$ versus $1/m$ with $N=m$ for $10\le m \le 200$. Constant step sizes. \mbox{Do-IT}: dark bullets, \mbox{CS-IT}: dark squares, \mbox{MCS-IT}: dark stars, \mbox{HV-IT}: dark diamonds.}
\label{Fig2Dbutcon_ADI}
\end{figure}

As a third example we select an American butterfly option on the maximum of two asset prices with payoff 
\[
\phi(s_1,s_2) = \max(s-K_1,0)-2\max(s-K,0)+\max(s-K_2,0)
\] 
with $s=\max(s_1,s_2)$ and $K=(K_1+K_2)/2$.
For this option the early exercise region encompasses all points $(s_1,K)$ and $(K,s_2)$ with $0\le s_1, s_2 \le K$.
We choose financial parameter values
\begin{equation}\label{2Dbutpar}
r=0.05,~~\sigma_1 = \sigma_2 = 0.30,~~\rho = 0.50,~~T=0.5,~~K_1=32,~~K_2=48.
\end{equation}
The obtained temporal errors for the $\theta$-based methods and the \mbox{ADI-IT} methods are displayed in Figures~\ref{Fig2Dbutcon_theta} and \ref{Fig2Dbutcon_ADI}, respectively.
The outcome is quite distinct from, and less favourable than, that in all foregoing (one- and two-asset) American option examples.
For the \mbox{BE-IT}, \mbox{BE-P} and \mbox{CN-P} methods a neat first-order convergence is observed.
Of these, the \mbox{CN-P} method is by far the most accurate.
For all other methods the converge behaviour is unclear.
We note that setting the correlation $\rho=0$, or computing the reference solution by the PR method instead of the \mbox{CN-P} method, does not change this conclusion.
Subsequent experiments for the two-asset American butterfly option up to the value $m=500$ suggest that for the \mbox{CN-IT} and PR methods there is convergence in $m=N$ of order $0.5$.
The converge behaviour of the \mbox{ADI-IT} methods is difficult to assimilate in this example and further research is required.

Employing the temporal grid points \eqref{variablestep}, corresponding to variable step sizes, leads in general again to a substantial improvement in accuracy for the \mbox{CN-P} method.
In the case of the two-asset butterfly option, a smooth second-order convergence behaviour is observed.
For the two-asset put options above, such a favourable result is obtained when the region of interest for the temporal error \eqref{temp_error} does not intersect with the early exercise boundary.

\section{Conclusions}\label{conclusions}
In this paper an ample numerical study has been performed for a collection of contemporary temporal discretization methods for PDCPs modelling the fair values of one- and two-asset American-style options.
To this purpose, a detailed numerical investigation has been carried out of the temporal discretization error \eqref{temp_error}.
Here the maximum norm is considered and the number of time steps $N$ has been taken directly proportional to the number of mesh points $m$ in each spatial direction.

Five American-style options are chosen for the numerical experiments: the one-asset put, the one-asset butterfly, the two-asset put on the minimum, the two-asset put on the arithmetic average, and the two-asset butterfly on the maximum.
For the temporal discretization, the backward Euler (BE) and Crank--Nicolson (CN) methods are selected together with three ADI schemes: Douglas (Do), Modified Craig--Sneyd (MCS) and Hundsdorfer--Verwer (HV).
For the numerical treatment of the LCPs that occur in each time step, the explicit payoff (EP) approach, the Ikonen--Toivanen (IT) splitting approach and the penalty (P) approach are considered.  
In addition to this, the Peaceman--Rachford (PR) method has been selected, which is related to the CN-IT method.
For the ADI schemes, only the combination with the IT splitting approach is studied in the present paper.

The two explicit payoff methods, \mbox{BE-EP} and \mbox{CN-EP}, have been considered just for one-asset options.
They show a temporal convergence order equal to 1.0 for the one-asset put option, but in the case of the one-asset butterfly option their temporal errors turn out to be large and convergence appears to be slow.

In contrast, for all five options above, the \mbox{BE-IT} and \mbox{BE-P} methods always show a temporal convergence order close to 1.0 and the \mbox{CN-P} method a convergence order between 1.0 and 1.3.
By employing suitable variable step sizes, defining the temporal grid points \eqref{variablestep}, the \mbox{CN-P} method reveals a favourable convergence order close to 2.0 whenever the early exercise boundary is not contained in the region of interest.
For all other methods under consideration, using these variable step sizes does unfortunately not lead to an improvement in their convergence behaviour.

The \mbox{CN-IT} and PR methods always show a convergence order between approximately 1.0 and 1.3, except for the two-asset butterfly option, where it appears to reduce to 0.5.

Concerning the \mbox{ADI-IT} methods, the \mbox{Do-IT} method with $\theta=1/2$ shows a convergence order about equal to 1.0 for both two-asset put options.
The \mbox{MCS-IT} methods with $\theta=1/3$ and $\theta=1/2$ and the \mbox{HV-IT} method with $\theta=1/(2+\sqrt{2})$ show convergence orders approximately equal to 1.3 and 1.0 for these two options, respectively.
The convergence behaviour of the \mbox{ADI-IT} methods is unclear in the case of the two-asset butterfly option.

The above observations on the temporal convergence behaviour of the methods employing the IT splitting approach appear to be largely new in the literature.
Only for the BE-IT method a directly related theoretical result is known to us, see \cite{KR_HH15}.
The numerical results in this paper are in agreement with this theoretical result.
For the methods using the penalty approach, our observations agree well with the (theoretical and practical) findings in \cite{KR_FV02}.
Clearly, a temporal convergence order close to or equal to two is only observed in the experiments in this paper for the \mbox{CN-P} method applied with suitable variable step sizes.

Comparing the size of the temporal errors of the different methods with constant step size, the experiments suggest that for the two methods \mbox{BE-IT}, \mbox{BE-P} these are always very similar, and the same is valid for the three methods \mbox{CN-IT}, \mbox{CN-P}, PR.
The latter group was found to be always significantly more accurate than the former group.
Further, the PR method proved to be often somewhat more accurate than the \mbox{CN-IT} method.
The ADI-based methods \mbox{MCS-IT} and~\mbox{HV-IT} revealed a similar accuracy to the CN-based methods in the case of the two-asset put options.
The \mbox{Do-IT} method was significantly less accurate than these.
With the pertinent variable step sizes, the \mbox{CN-P} method has been found to be the most accurate in general among all methods under consideration.

Based on the numerical experiments discussed in this paper, and taking into account the amount of computational work per time step, the  \mbox{MCS-IT} and~\mbox{HV-IT} methods are recommended in the numerical solution of \eqref{PDCP} for two-asset American-style options whenever the payoff function and the financial parameters are standard.
If the payoff function is more advanced, such as the (nonconvex) two-asset butterfly on the maximum, we recommend the \mbox{CN-P} method with variable step sizes.
The \mbox{BE-IT} and \mbox{BE-P} methods are advocated if general applicability is important, which goes at the expense of temporal accuracy.
As a comprise between general applicability and temporal accuracy, the PR method forms a good candidate.

\section*{Acknowledgements}
This work has been supported by the European Union in the FP7-PEOPLE-2012-ITN Program under Grant Agreement Number 304617 (FP7 Marie Curie Action, Project Multi-ITN STRIKE - Novel Methods in Computational Finance).
\clearpage

\clearpage

\clearpage

\clearpage

\end{document}